\documentclass{emulateapj}
\usepackage{longtable, array}

\begin{document}

\setlength\LTcapwidth{\textwidth}

\newcommand\blfootnote[1]{%
  \begingroup
  \renewcommand\thefootnote{}\footnote{#1}%
  \addtocounter{footnote}{-1}%
  \endgroup
}

\title{Probability of Physical Association of 104 Blended Companions to \textit{Kepler} Objects of Interest Using Visible and Near-Infrared Adaptive Optics Photometry}
\author{Dani Atkinson\altaffilmark{1}, Christoph Baranec\altaffilmark{1}, Carl Ziegler\altaffilmark{2}, Nicholas Law\altaffilmark{2}, Reed Riddle\altaffilmark{3}, and Tim Morton\altaffilmark{4}}
\altaffiltext{1}{Institute for Astronomy, University of Hawai`i at M\={a}noa, Hilo, Hawai`i 96720-2700, USA}
\altaffiltext{2}{Division of Physics and Astronomy, University of North Carolina at Chapel Hill, Chapel Hill, NC 27599-3255, USA}
\altaffiltext{3}{Division of Physics, Mathematics, and Astronomy, California Institute of Technology, Pasadena, CA 91125, USA}
\altaffiltext{4}{Department of Astrophysical Sciences, Princeton University, Princeton, NJ 08544, USA}
\blfootnote{Please direct correspondence to atkinson@ifa.hawaii.edu}

\begin{abstract}
We determine probabilities of physical association for stars in blended Kepler Objects of Interest, and find that $14.5\%^{+3.8\%}_{-3.4\%}$ of companions within $\sim4\arcsec$ are consistent with being physically unassociated with their primary. This produces a better understanding of potential false positives in the Kepler catalog and will guide models of planet formation in binary systems. Physical association is determined through two methods of calculating multi-band photometric parallax using visible and near-infrared adaptive optics observation of 84 KOI systems with 104 contaminating companions within $\sim4\arcsec$. We find no evidence that KOI companions with separation of less than $1\arcsec$ are more likely to be physically associated than KOI companions generally. We also reinterpret transit depths for 94 planet candidates, and calculate that $2.6\% \pm 0.4\%$ of transits have $R > 15R_{\earth}$, which is consistent with prior modeling work.
\end{abstract}


\section{Introduction}
The \textit{Kepler} mission had a simple observing strategy: it observed a 105 deg$^2$ field in Cygnus near-continuously with an unfiltered wideband camera. Its main data output were the light curves of target stars, in which it found transits and measured their depth and timing. The conversion of this transit information to planetary characteristics requires the stellar parameters of the host, which the \textit{Kepler} telescope could not provide itself. Stellar characterization is then dependent on data from other sources, typically photometric observations performed for the Kepler Input Catalog in the visible and by 2MASS in the near-infrared \citep{brown2011,liebert1995,huber2014}.

A complication arises from the vulnerability of \textit{Kepler}'s relatively large 4\arcsec pixels to the misinterpretation of unresolved binaries as single stars \citep{borucki2010}. These unseen companions dilute the transit by making it appear shallower relative to its host star, and thus the transiting object's size is underestimated. Photometric characterization of the host star is also distorted by the blended light.

Many of these blended and contaminating companions can be identified in the \textit{Kepler} data by careful examination of the light curve data for irregularities, including secondary transits (indicative of an eclipsing binary) and shifts in the star's centroid coincident with observed transits \citep{batalha2010,tenenbaum2013}. These techniques have proven largely successful in screening out many false positives, and though it has been shown that of the remaining contaminated KOIs the great majority ($>90\%$) are not false positives, many transiting planets are still larger than interpreted \citep{morton2011,fressin2013,santerne2013,ciardi2015,desert2015}. Further validation then requires finding contaminating companions either indirectly, e.g. with transit photometry \citep{colon2012,colon2015} or directly, e.g. high angular resolution imaging \citep{morton2012}. The necessary sub-arcsecond resolution to find these contaminating companions can be achieved on the ground by several techniques, most notably lucky/speckle imaging \citep{horch2012,lillo-box2012,lillo-box2014} and adaptive optics \citep{adams2012, adams2013, dressing2014}.

With 6395 Kepler Objects of Interest (KOI) to vet \citep{coughlin2015}, we adopted a strategy to conduct a comprehensive survey with Palomar 1.5-m/Robo-AO \citep{baranec2013b,baranec2013a} and use Keck II/NIRC2 to follow up on secure and likely detections of companions. To date we have reported the optical detection of 53 contaminating companions within $2.5\arcsec$ in a sample of 715 KOIs \citep{law2014} and 206 companions within $\sim4\arcsec$ to 1629 KOIs \citep{baranec2016, ziegler2016}. The addition of near-infrared observations to the existing visible data permits us to perform characterization of the detected stars, estimate their photometric parallax and the likelihood of physical association between primary/companion pairs, and calculate reinterpreted sizes for individual planet candidates. In several cases we have also found additional unseen companions.

Section~\ref{observations} of this paper describes the observations made and the image reduction process for Keck II/NIRC2 data. Section~\ref{analysis} describes the derivation of photometric and stellar characteristics from the objects studied, including techniques used for fitting to stellar type and results thereof. Section~\ref{discussion} discusses the spectral fit results in context of the entire KOI catalog. The paper concludes in Section~\ref{conclusions} with an overview of our findings and an outline of future avenues of investigation.

\section{Observations and Data Reduction}\label{observations}

   \begin{figure}
   \begin{center}
   \includegraphics[height = 3.9cm]{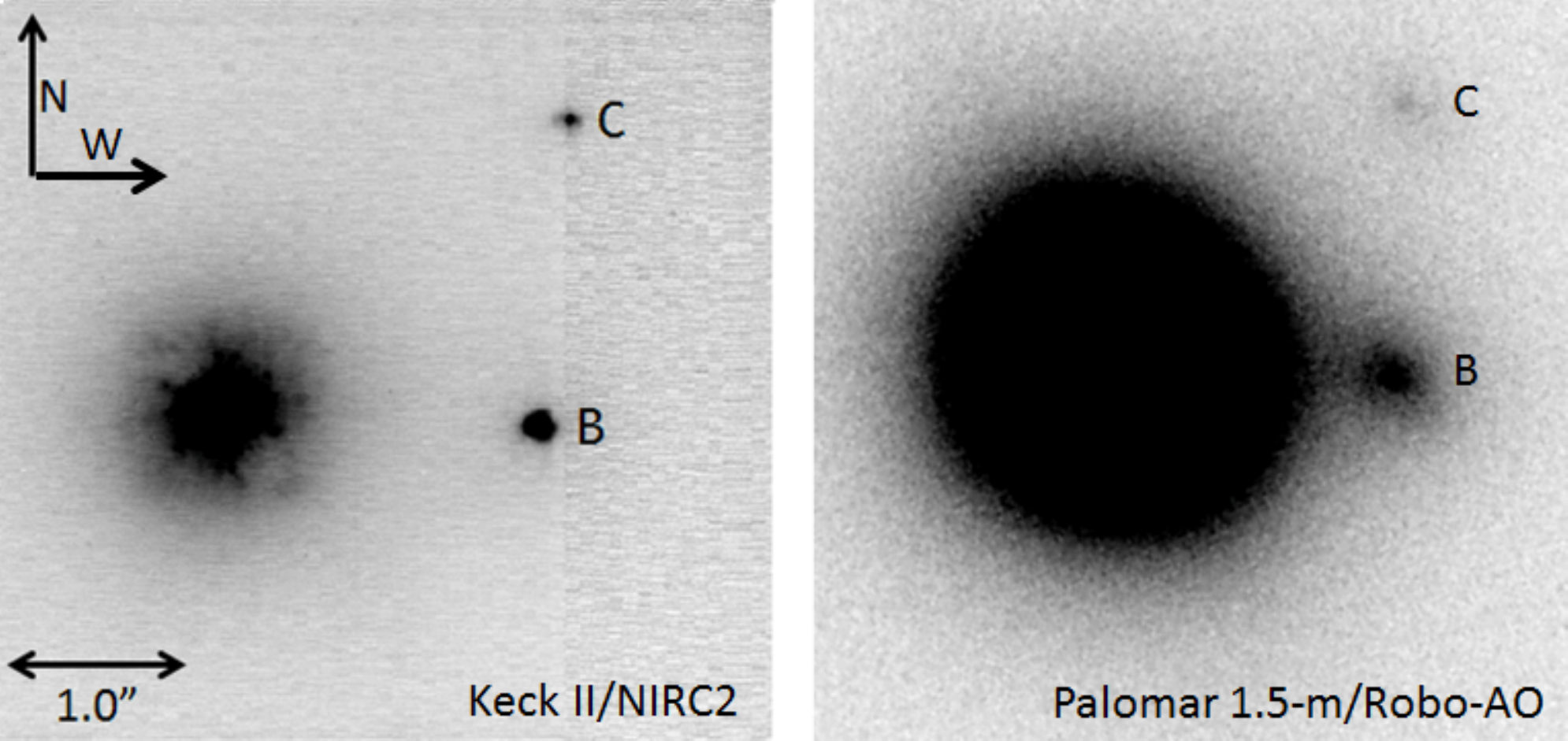}
   \end{center}
   \caption[example]{ \label{fig:example} 
Images of KOI-268 from both Keck II/NIRC2 (\textit{left}) and Palomar 1.5-m/Robo-AO, presented as an example. Visible in both images are companions B and C at separations from A of 1\farcs75 and 2\farcs53, respectively.}
   \end{figure} 

The initial observations identifying companion candidates are from multiple Robo-AO observing runs on the Palomar Observatory 1.5-m telescope, spanning July to September 2012, April to October 2013, June to September 2014, and June 2015. Observations were in either Sloan-\textit{i} or a long-pass 600nm (LP600) filter, the latter being similar to the \textit{Kepler}-bandpass when combined with the EMCCD's quantum efficiency curve for red/cool stars. Images were automatically reduced by the Robo-AO observing pipeline \citep{law2014}.

The near-infrared observations are from the NIRC2 instrument on the 10-m Keck II telescope, conducted 2013 June 24, August 24 and 25, 2014 August 17, 2015 July 25, and August 4 in the \textit{J}, \textit{H}, \textit{K}, and/or $K_p$ filters in the narrow mode of NIRC2 (9.952 mas pixel$^{-1}$; \citealt{Yelda2010}). For KOIs brighter than $m_V \sim 13$ we typically used the KOI as the guide star in natural guide star mode, and for fainter KOIs we used the laser guide star, with the KOI as the tip-tilt-focus guide star \citep{wizinowich2006, vandam2006}. An initial 30s exposure was taken for each target, and we waited for the low-bandwidth wavefront sensor to settle if the laser was used. The integration time and number of coadds per detector readout were adjusted to keep the peak of the stellar PSF counts less than 8,000 ADU per single integration (roughly half the dynamic range where sensitivity of the detector is linear), while maintaining a total exposure time of 30s. Dithered images were then acquired with the primary centered in the 3 lowest noise quadrants using the `bxy3 2.5' command, for a total exposure time of 90s. Occasional dither failures, particularly on 4 Aug 2015, resulted in exposures where the target is centered on the detector.

Images are first sky-subtracted and then flat-fielded. A pipeline developed for this investigation is then used to automatically pick out companion stars by spatially binning pixels and selecting the locations of the brightest bins as candidates. These candidates then have their radius measured in the eight cardinal and intermediate directions from their local centroid. This measurement steps in the given direction until it finds a value consistent with the measured background value within a specified confidence interval (initially $3\sigma$). If the median radius is both larger than a specified cutoff value (typically 3 pixels) and the standard deviation of the measured radii smaller than the same cutoff, the candidate is accepted as a star. The brightest star in the field is assumed to be the primary unless manually specified otherwise (the narrow $10\arcsec$ NIRC2 field makes this a rare occurrence). If a companion is not found, the background confidence interval and radius cutoff are adjusted to optimize for close ($<0\farcs5$) companions and the procedure repeated. If the procedure fails to find a companion, or if it finds multiple companions, a warning message notifies the operator to review the source image. 

For the majority of targets the pipeline correctly locates the primary and any present companions, but manual validation is necessary for many targets largely because the speckles in the point spread function (PSF) are mistaken for stars.  To avoid this misinterpretation the pipeline cross-references stars found in different filters for the same object, and discards any objects that do not appear in multiple filters. In some cases, images were only taken with one filter (typically $K_{p}$), and thus cross-referencing is not possible. These targets are then manually vetted and removed if visually confirmed to be associated with the primary PSF (i.e. coincident with rays projecting from the primary and presenting a PSF inconsistent with other imaged stars).

The separation and phase angle of each companion are measured from these individual reduced images, with uncertainties measured from the variability in measurements across all available images, and corrected for distortion using the most recent solution \citep{service2016}. Last, images are co-added into a single composite image for each target and filter for use in photometry.

\subsection{Aperture Photometry}
For the majority of our Keck data, the diffraction-limited resolution makes simple aperture photometry sufficient for measuring the contrast between the two stars. To account for the overlap of the stars' PSF envelopes, a matching aperture from the location opposite each star relative to its companion is used to estimate background subtraction, if available. In cases where the corresponding Robo-AO results were unavailable the method was also applied to those images, using the known position of the companion taken from the Keck analysis.

Systematic error from aperture size is our primary source of uncertainty, and is measured as the standard deviation of contrast across a range of aperture sizes from 1 to 3 FWHM in 0.5 FWHM increments. Injected companions are used to estimate further uncertainties typically yielding an error of $5\%$.

\section{Analysis}\label{analysis}
\subsection{Photometric Classification of Stars}\label{fitting}
By combining our contrast measurements with extant \textit{JHK} photometry for the blended system (from the Exoplanet Archive\footnote{http://exoplanetarchive.ipac.caltech.edu Most \textit{JHK} magnitudes are from the 2MASS catalog \citep{liebert1995}.}) we derive the multi-color photometry for all components of each system. For blended magnitudes lacking a reported uncertainty (\textit{i},\textit{Kep}), one was estimated based on the measurement's reported source as recommended by the guide supplied by MAST \citep{mikulski}.\footnote{The documentation on Kepler magnitude sources at archive.stsci.edu/kepler/help/columns.html under heading `Kepmag Source' describes the respective uncertainties for the Kepler magnitude sources.}

Multi-color photometry allows characterization of the stars, necessary as the existing data on these objects is drawn from a blended target. Effective temperature is relatively strongly correlated with color-color photometry, but stellar radius (upon which our measurement of a transiting planet's size is dependent) exhibits a much weaker correlation for late-type stars. To demonstrate the systematic biases inherent in photometric type-fitting, we present fitting results from 2 different photometric datasets.

The first dataset is a set of metallicity- and age-agnostic stellar SEDs, originally assembled from a heterogenous set of models and data for an investigation of the Praesepe and Coma Berenices open clusters by \citet{kraus2007}, henceforth KH07, and which has previously been used for photometric fitting of exoplanet host stars (e.g. \citealt{bechter2014,wang2014,wang2015}). Photometric values for the \textit{Kepler}-band were computed by the method described in \citet{brown2011} using an arithmetic combination of \textit{gri} colors. The list of types and magnitudes from KH07 is expanded with missing types linearly interpolated from existing data, and an additional 9 intermediate types also interpolated between each two adjacent integer stellar types (e.g. type G2.5 is linearly halfway between G2 and G3). This makes a table of 521 entries from B8 to L0 to be fitted to as standards, and with matching absolute magnitudes and radii. The interpolated decimal types are not reported directly but those entries are used to refine radius/distance estimates. Radii for spectral types are drawn from \citet{habets1981}.

To fit types we use a Monte Carlo technique, generating a Gaussian distribution for each of the photometric combinations $J-K$, $H-K$, $i-K$, and $Kep-K$, if information in the respective filters is present. $K$ was chosen as the baseline as it produces the most precise contrast measurements and occupies the longest wavelength. Extinction is corrected for during this fitting process, relying on the canonical $A_V$ for each target in the Kepler catalog and adjusted for the various filters/bandpasses via the standard relations from \citet{cardelli1989}: $A_{Kep} = 0.896A_V$, $A_i = 0.321A_V$, $A_J = 0.158A_V$, $A_H = 0.100A_V$, $A_K = 0.060A_V$.

Each time beginning in the center of the list of standards, the Gaussian-generated photometry is compared to each canonical type's set of magnitudes to measure the error for all available data, as
\begin{equation}
R_{f1-f2} = (m_{f1,*} - m_{f2,*}) - (M_{f1,std} - M_{f2,std})
\end{equation}
where $m_{f\#,*}$ and $M_{f\#,typ}$ are the star's measured apparent magnitude and the standard's absolute magnitude in filters $f1$ and $f2$. The quality of the fit against the given standard is then
\begin{equation}
R^2 = \sum\limits_{J,H,i,Kep} \frac{R^2_{filter-K}}{w^2_{filter-K}}
\label{rsquared}
\end{equation}
where $w_{filter-K}$ is the weight of the respective filter combination as
\begin{equation}
w_{filter-K} = \sqrt{\sigma^2_{filter} + \sigma^2_K}
\end{equation}
Note that Equation~\ref{rsquared} does not require normalization as the number of filters used is consistent for a given star. $R^2$ is also measured for comparisons to standards in both directions (earlier- and later-type), and moved if a lower $R^2$ is found in either. The process repeats until it finds a minimum $R^2$. The type, radius, and absolute magnitudes are recorded and the next member of the Gaussian-generated list is fit to the standards in the same way. After fitting every entry on the list, the means of type, radius, and absolute magnitudes are taken as the fitted values, and the standard deviation of the latter two are their uncertainties.

The second set is the \textit{Kepler} Input Catalog's (KIC) primary standard stars as reported in \citet{brown2011}, henceforth B11. The advantage of this catalog is that these stars are in the \textit{Kepler} field and therefore reasonably representative of stars in our sample. The subset of standard stars with which each studied KOI component's photometry is consistent to $1\sigma$ was used to compute a mean and standard deviation for the star's stellar parameters. Subsets are typically $10\%$ of the full list of primary standard stars (or $\sim30$ stars) for each fitted object. Given the relatively poor correlation in the KIC standards of any of the measured stellar characteristics with NIR-only color comparisons, only $Kep - K$ and $i - K$ measurements were used. For stars without available $Kep$- or $i$-band measurements (with LP600 approximating $Kep$), a fit is not produced. If the photometry for a target fits two or fewer primary standard stars, its results are omitted. The effective temperature is then fit to a modeled stellar catalog to produce absolute magnitudes \citep{pecaut2013}, and compared to apparent magnitudes in turn to estimate distance. The KIC standard magnitudes were corrected for extinction/reddening as calculated from \citet{schlafly2011}.

The fitted values for both methods are displayed in Table~\ref{stellarparams}. The two methods are in broad agreement, although disparities are apparent for M-type companions in particular as the KIC primary standards contain few objects in that temperature range and exhibit a systemic overestimation of late-K and M-type radii that is not corrected for here \citep{muirhead2012}. While the range of potential fit types for KH07 covered the full main sequence from B8 to M9, almost all stars fit to late types F-M. 

Notably, two stars are presented as type B8, and five are too poorly restrained to produce KH07 fits. For the former, B8 is the end of the fitting range, and indicates they are too blue (adjusted for reddening) to fit to our list of main sequence stars. These stars are then very distant O/B types, and at 2 of 159 total stars analyzed by KH07 make up $\sim1\%$ of the sample. As we are unable to fit above B8 their properties are not well constrained, and we present radius and distance as lower limits.

KH07 fitting fails to fit five stars in the sample due to relatively poor photometric constraints in one or more observation bands.

\subsection{Uncertainties of Fitted Characteristics}
As described above, the photometric uncertainties arise largely from sampling the contrast for a range of photometric apertures and the inherent 5\% error measured by injection tests. For the stellar type fitting to KH07, the full Monte Carlo fit measures the uncertainty of derived characteristics. Gaussian distributions matching each apparent magnitude measurement and uncertainty thereof are grouped into sets, and each set has its type and other fitted characteristics calculated by the method described above. The measured uncertainty in the derived characteristics is then the standard deviation of the full set of measurements. Uncertainties from the B11 fits are simply the measured standard deviations of the respective parameters of $1\sigma$ consistent KIC standard stars. Uncertainties may be underestimated due to the granularity of the data being fit.

B11 also reports that use of photometric fitting on \textit{Kepler} primary standard stars results in uncertainties of approximately 200K for effective temperature and 0.2 dex for stellar radius without prior constraints on stellar age or metallicity. We take this to be generally applicable to all our photometric type-fitting, but it is not factored into the uncertainties reported in Table~\ref{stellarparams}.

\begin{longtable*}{c c c c c c c c c}
\setlength{\extrarowheight}{3pt}\\
\caption{Fitted Stellar Parameters}\label{stellarparams}\\
& \multicolumn{4}{|c}{via \citet{kraus2007}} & \multicolumn{4}{|c}{via \citet{brown2011}}\\
\hline
object & $SpT$ & $R/R_{Sun}$ & $dist (pc)$ & $\sigma_{unassoc}$ & $T_{eff}$ & $R/R_{Sun}$ & $dist (pc)$ & $\sigma_{unassoc}$\\
\endfirsthead
\caption{(continued)}\\
 & \multicolumn{4}{|c}{via \citet{kraus2007}}  & \multicolumn{4}{|c}{via \citet{brown2011}}\\
\hline
object & $SpT$ & $R (R_{Sun})$ & $dist (pc)$ & $\sigma_{unassoc}$ & $T_{eff}$ & $R (R_{Sun})$ & $dist (pc)$ & $\sigma_{unassoc}$\\
\endhead
\endfoot

\caption{Stellar parameters as a result of two different fitting techniques. $\sigma_{unassoc}$ is the certainty (in standard deviations) that each companion is physically unassociated with its host due to their respective distances. The Kraus \& Hillenbrand fit yields a stellar type and corresponding radius \citep{kraus2007}. Values are interpolated between the table items in the source for improved precision. The fit to the KIC primary standards from Brown yields effective temperature and stellar radius for all stars with sufficient photometry, as produced by comparison to stars with similar color-color measurements among the 279 entries in the KIC Primary Standard catalog \citep{brown2011}. As noted, photometric type-fitting of \textit{Kepler} targets has been found to have a limiting accuracy of $\pm$200K and $\sim$0.2dex respectively, which is largely a function of age/composition and is not taken into account here. For each primary/companion pair a distance measurement was produced from the measured apparent and fitted absolute magnitudes, and used to generate a confidence of non-association between the two objects.}
\endlastfoot
0190A & G0 & $1.18^{+0.04}_{-0.07}$ &  $921^{+127}_{-109}$ &      &  $6041^{+116}_{-126}$ & $1.04^{+0.03}_{-0.03}$  & $1019^{+52}_{-93}$ &     \\
0190B & K3 & $0.93^{+0.03}_{-0.02}$ &  $716^{+89}_{-65}$   & 1.46 &  $4904^{+345}_{-265}$ & $0.83^{+0.13}_{-0.07}$  & $779^{+122}_{-77}$ & 1.56\\
0191A & G0 & $1.07^{+0.05}_{-0.03}$ &  $934^{+82}_{-69}$   &      &  $5855^{+109}_{-184}$ & $1.00^{+0.02}_{-0.03}$  & $1115^{+93}_{-115}$ &     \\
0191B & F4 & $1.01^{+0.14}_{-0.07}$ & $2804^{+1007}_{-492}$& 3.75 &  $5110^{+378}_{-472}$ & $0.94^{+0.09}_{-0.12}$  & $2573^{+651}_{-386}$ & 3.67\\
0268A & F3 & $1.29^{+0.06}_{-0.06}$ &  $258^{+33}_{-34}$   &      &  $6136^{+143}_{-111}$ & $1.04^{+0.04}_{-0.03}$  & $230^{+18}_{-12}$ &     \\
0268B & K4 & $0.85^{+0.04}_{-0.06}$ &  $315^{+33}_{-33}$   & 1.22 &  $4807^{+423}_{-283}$ & $0.79^{+0.16}_{-0.06}$  & $392^{+81}_{-41}$ & 3.62\\
0268C & K3 & $0.95^{+0.31}_{-0.08}$ &  $904^{+3016}_{-290}$& 2.21 &  $4889^{+660}_{-352}$ & $0.82^{+0.28}_{-0.08}$  & $829^{+354}_{-133}$ & 4.46\\
0401A & G4 & $1.05^{+0.02}_{-0.02}$ &  $589^{+31}_{-44}$   &      &  $5890^{+152}_{-206}$ & $1.01^{+0.02}_{-0.03}$  & $744^{+70}_{-96}$ &     \\
0401B & K7 & $0.79^{+0.05}_{-0.08}$ &  $690^{+74}_{-78}$   & 1.20 &  $5000^{+970}_{-554}$ & $0.88^{+0.33}_{-0.17}$  & $1185^{+716}_{-315}$ & 1.37\\
0425A & F4 & $1.22^{+0.05}_{-0.04}$ & $1374^{+212}_{-138}$ &      &  $6088^{+113}_{-165}$ & $1.02^{+0.04}_{-0.03}$  & $1425^{+95}_{-115}$ &     \\
0425B & F2 & $1.20^{+0.06}_{-0.07}$ & $1947^{+343}_{-253}$ & 1.74 &  $6013^{+116}_{-137}$ & $1.03^{+0.03}_{-0.03}$  & $2024^{+122}_{-177}$ & 2.98\\
0511A & F3 & $1.25^{+0.02}_{-0.04}$ & $1067^{+77}_{-118}$  &      &  $6128^{+73}_{-112}$  & $1.04^{+0.03}_{-0.03}$  & $1061^{+38}_{-59}$ &     \\
0511B & K4 & $0.84^{+0.04}_{-0.07}$ &  $995^{+109}_{-115}$ & 0.45 &  $4796^{+456}_{-303}$ & $0.79^{+0.18}_{-0.06}$  & $1280^{+290}_{-149}$ & 1.42\\
0511C & K6 & $0.62^{+0.13}_{-0.16}$ & $2345^{+491}_{-609}$ & 2.08 &                       &                         &        \\
0628A & F5 & $1.24^{+0.02}_{-0.02}$ &  $876^{+40}_{-46}$   &      &  $6085^{+113}_{-192}$ & $1.04^{+0.04}_{-0.03}$  & $856^{+51}_{-104}$ &     \\
0628B & K9 & $0.64^{+0.08}_{-0.11}$ & $1070^{+140}_{-180}$ & 1.05 &                       &                         &        \\
0628C & K1 & $0.85^{+0.05}_{-0.07}$ & $2247^{+254}_{-286}$ & 4.75 &  $4744^{+533}_{-329}$ & $0.74^{+0.21}_{-0.07}$  & $2809^{+820}_{-385}$ & 5.03\\
0687A & F6 & $1.24^{+0.02}_{-0.04}$ &  $837^{+56}_{-72}$   &      &  $6012^{+119}_{-158}$ & $1.02^{+0.04}_{-0.03}$  & $806^{+55}_{-76}$ &     \\
0687B & K4 & $0.89^{+0.05}_{-0.07}$ &  $699^{+116}_{-83}$  & 1.01 &  $5486^{+417}_{-300}$ & $1.02^{+0.15}_{-0.07}$  & $1114^{+311}_{-176}$ & 1.67\\
0688A & F4 & $1.30^{+0.01}_{-0.01}$ & $1204^{+49}_{-54}$   &      &  $6158^{+86}_{-158}$  & $1.03^{+0.02}_{-0.02}$  & $1057^{+50}_{-87}$ &     \\
0688B & K0 & $0.99^{+0.07}_{-0.04}$ & $1166^{+239}_{-125}$ & 0.16 &  $5122^{+342}_{-342}$ & $0.87^{+0.09}_{-0.09}$  & $1174^{+249}_{-142}$ & 0.78\\
0712A & F1 & $1.26^{+0.04}_{-0.17}$ & $1100^{+124}_{-221}$ &      &  $6228^{+78}_{-169}$  & $1.07^{+0.03}_{-0.03}$  & $1045^{+52}_{-76}$ &     \\
0712B & K1 & $0.95^{+0.05}_{-0.03}$ &  $637^{+82}_{-54}$   & 1.96 &  $5204^{+472}_{-318}$ & $0.89^{+0.16}_{-0.09}$  & $762^{+220}_{-118}$ & 1.22\\
0931A & F3 & $1.27^{+0.04}_{-0.04}$ & $1720^{+195}_{-186}$ &      &  $6140^{+74}_{-106}$  & $1.05^{+0.03}_{-0.04}$  & $1654^{+57}_{-94}$ &     \\
0931B &    &                        &                      &      &  $5943^{+139}_{-225}$ & $1.01^{+0.04}_{-0.03}$  & $6491^{+602}_{-817}$ & 5.91\\
0984A & G5 & $1.03^{+0.03}_{-0.02}$ &  $267^{+21}_{-19}$   &      &  $5806^{+191}_{-304}$ & $1.00^{+0.05}_{-0.06}$  & $326^{+48}_{-44}$ &     \\
0984B & G5 & $1.03^{+0.03}_{-0.02}$ &  $273^{+22}_{-19}$   & 0.21 &  $5717^{+185}_{-302}$ & $0.99^{+0.04}_{-0.07}$  & $317^{+44}_{-46}$ & 0.14\\
0987A & G7 & $1.01^{+0.01}_{-0.01}$ &  $270^{+13}_{-6}$    &      &  $5656^{+241}_{-333}$ & $0.99^{+0.08}_{-0.08}$  & $325^{+55}_{-49}$ &     \\
0987B & M0 & $0.74^{+0.05}_{-0.05}$ &  $458^{+47}_{-37}$   & 4.79 &                       &                         &        \\
1066A & F6 & $1.13^{+0.05}_{-0.05}$ & $1342^{+152}_{-108}$ &      &  $5932^{+102}_{-157}$ & $1.02^{+0.02}_{-0.03}$  & $1535^{+110}_{-145}$ &     \\
1066B & G9 & $0.85^{+0.03}_{-0.04}$ & $2894^{+245}_{-259}$ & 5.17 &  $4608^{+256}_{-285}$ & $0.74^{+0.06}_{-0.05}$  & $3333^{+289}_{-419}$ & 4.15\\
1067A & F3 & $1.29^{+0.01}_{-0.01}$ & $1431^{+60}_{-65}$   &      &  $6146^{+80}_{-205}$  & $1.04^{+0.03}_{-0.03}$  & $1280^{+72}_{-125}$ &     \\
1067B & K2 & $0.89^{+0.03}_{-0.02}$ & $2103^{+214}_{-175}$ & 3.63 &  $4833^{+409}_{-282}$ & $0.80^{+0.15}_{-0.06}$  & $2520^{+493}_{-269}$ & 4.45\\
1112A & F5 & $1.22^{+0.04}_{-0.06}$ & $1140^{+141}_{-160}$ &      &  $6160^{+88}_{-87}$   & $1.07^{+0.04}_{-0.03}$  & $1232^{+63}_{-43}$ &     \\
1112B & K8 & $0.73^{+0.04}_{-0.04}$ & $1528^{+120}_{-104}$ & 2.21 &                       &                         &        \\
1151A & F9 & $1.10^{+0.08}_{-0.05}$ &  $501^{+76}_{-44}$   &      &  $5776^{+148}_{-270}$ & $0.99^{+0.02}_{-0.06}$  & $534^{+61}_{-71}$ &     \\
1151B & K8 & $0.71^{+0.19}_{-0.30}$ &  $738^{+373}_{-289}$ & 0.79 &  $4620^{+488}_{-418}$ & $0.72^{+0.20}_{-0.07}$  & $991^{+240}_{-177}$ & 2.44\\
1214A & G8 & $1.00^{+0.02}_{-0.02}$ &  $656^{+47}_{-26}$   &      &  $5600^{+256}_{-230}$ & $0.99^{+0.08}_{-0.07}$  & $836^{+136}_{-102}$ &     \\
1214B & B8 & $\geq1.5$              & $\geq5591$           & 7.29 &                       &                         &        \\
1274A & G7 & $1.00^{+0.01}_{-0.01}$ &  $363^{+14}_{-9}$    &      &  $5602^{+250}_{-319}$ & $0.97^{+0.09}_{-0.08}$  & $443^{+76}_{-65}$ &     \\
1274B & K9 & $0.63^{+0.17}_{-0.26}$ &  $585^{+191}_{-254}$ & 0.87 &                       &                         &        \\
1375A & F5 & $1.25^{+0.01}_{-0.02}$ &  $809^{+28}_{-41}$   &      &  $6136^{+89}_{-103}$  & $1.05^{+0.04}_{-0.03}$  & $811^{+31}_{-47}$ &     \\
1375B & K3 & $0.88^{+0.21}_{-0.19}$ & $1862^{+1419}_{-480}$& 2.19 &  $5097^{+433}_{-375}$ & $0.86^{+0.16}_{-0.09}$  & $2305^{+609}_{-325}$ & 4.58\\
1442A & G4 & $1.08^{+0.05}_{-0.03}$ &  $322^{+26}_{-14}$   &      &  $5791^{+137}_{-294}$ & $0.99^{+0.02}_{-0.06}$  & $352^{+40}_{-50}$ &     \\
1442B & M7 & $0.30^{+0.16}_{-0.01}$ &  $350^{+65}_{-29}$   & 0.72 &                       &                         &        \\
1447A & A6 & $1.45^{+0.04}_{-0.03}$ & $1206^{+62}_{-54}$   &      &                       &                         &     \\
1447B & K3 & $0.91^{+0.03}_{-0.02}$ &  $517^{+43}_{-34}$   & 9.98 &  $4852^{+430}_{-216}$ & $0.77^{+0.17}_{-0.06}$  & $595^{+122}_{-59}$ &        \\
1536A & F5 & $1.27^{+0.01}_{-0.01}$ &  $550^{+20}_{-18}$   &      &  $6229^{+71}_{-89}$   & $1.08^{+0.03}_{-0.03}$  & $551^{+19}_{-25}$ &     \\
1536B & K3 & $0.80^{+0.24}_{-0.34}$ & $1614^{+1388}_{-600}$& 1.77 &  $5117^{+449}_{-439}$ & $0.87^{+0.15}_{-0.10}$  & $2169^{+607}_{-349}$ & 4.63\\
1546A & F2 & $1.22^{+0.05}_{-0.09}$ & $1259^{+177}_{-256}$ &      &  $6116^{+75}_{-124}$  & $1.05^{+0.03}_{-0.03}$  & $1319^{+55}_{-80}$ &     \\
1546B & K0 & $0.93^{+0.06}_{-0.04}$ &  $939^{+155}_{-98}$  & 1.07 &  $5111^{+384}_{-377}$ & $0.87^{+0.12}_{-0.10}$  & $1108^{+264}_{-148}$ & 0.76\\
1546C &    &                        &                      &      &  $5873^{+300}_{-429}$ & $1.00^{+0.06}_{-0.10}$  & $5685^{+1077}_{-1181}$ & 3.69\\
1546D & F7 & $0.91^{+0.10}_{-0.08}$ & $2527^{+798}_{-415}$ & 2.81 &  $5208^{+386}_{-492}$ & $0.92^{+0.09}_{-0.12}$  & $3175^{+834}_{-522}$ & 3.54\\
1613A & F5 & $1.27^{+0.05}_{-0.05}$ &  $404^{+60}_{-46}$   &      &                       &                         &     \\
1613B & G4 & $1.07^{+0.11}_{-0.09}$ &  $419^{+101}_{-89}$  & 0.14 &                       &                         &        \\
1700A & G8 & $0.97^{+0.02}_{-0.02}$ &  $602^{+36}_{-33}$   &      &  $5228^{+216}_{-261}$ & $0.89^{+0.06}_{-0.07}$  & $648^{+80}_{-68}$ &     \\
1700B & K3 & $0.89^{+0.03}_{-0.03}$ &  $632^{+52}_{-40}$   & 0.56 &  $4614^{+409}_{-272}$ & $0.74^{+0.15}_{-0.06}$  & $685^{+119}_{-75}$ & 0.34\\
1784A & F7 & $1.16^{+0.05}_{-0.06}$ &  $751^{+76}_{-73}$   &      &  $5939^{+100}_{-135}$ & $1.01^{+0.03}_{-0.03}$  & $817^{+54}_{-74}$ &     \\
1784B & F2 & $1.26^{+0.03}_{-0.06}$ & $1365^{+146}_{-176}$ & 3.20 &  $6113^{+85}_{-105}$  & $1.04^{+0.03}_{-0.03}$  & $1292^{+51}_{-68}$ & 5.47\\
1845A & K2 & $0.94^{+0.02}_{-0.01}$ &  $411^{+24}_{-18}$   &      &                       &                         &     \\
1845B & M4 & $0.46^{+0.29}_{-0.19}$ &  $651^{+445}_{-285}$ & 0.84 &                       &                         &        \\
1845C & A7 & $1.07^{+0.17}_{-0.10}$ &$4407^{+2110}_{-1089}$& 3.67 &                       &                         &        \\
1880A & K8 & $0.78^{+0.01}_{-0.06}$ &  $206^{+3}_{-17}$    &      &                       &                         &     \\
1880B & G7 & $0.93^{+0.07}_{-0.06}$ & $2094^{+492}_{-290}$ & 6.51 &  $4717^{+418}_{-334}$ & $0.76^{+0.16}_{-0.08}$  & $2046^{+365}_{-279}$ &        \\
1884A & F9 & $1.06^{+0.07}_{-0.04}$ & $1094^{+147}_{-123}$ &      &  $5890^{+159}_{-252}$ & $1.01^{+0.04}_{-0.05}$  & $1366^{+143}_{-193}$ &     \\
1884B & K8 & $0.63^{+0.17}_{-0.22}$ & $1381^{+449}_{-514}$ & 0.54 &                       &                         &        \\
1884C & M1 & $0.51^{+0.21}_{-0.15}$ & $1330^{+587}_{-424}$ & 0.53 &                       &                         &        \\
1884D & M2 & $0.41^{+0.17}_{-0.10}$ & $1335^{+642}_{-392}$ & 0.58 &                       &                         &        \\
1891A & K0 & $0.95^{+0.01}_{-0.02}$ &  $687^{+27}_{-46}$   &      &  $5417^{+248}_{-158}$ & $0.97^{+0.08}_{-0.06}$  & $872^{+118}_{-93}$ &     \\
1891B &    &                        &                      &      &  $5802^{+284}_{-402}$ & $1.00^{+0.06}_{-0.09}$  & $8742^{+1733}_{-1568}$ & 5.00\\
1916A & F2 & $1.31^{+0.01}_{-0.01}$ & $1046^{+43}_{-36}$   &      &                       &                         &     \\
1916B & K5 & $0.89^{+0.03}_{-0.04}$ &  $672^{+85}_{-59}$   & 4.05 &  $4556^{+266}_{-276}$ & $0.73^{+0.08}_{-0.03}$  & $685^{+65}_{-84}$ &        \\
1979A & F5 & $1.23^{+0.02}_{-0.07}$ &  $568^{+27}_{-59}$   &      &  $6012^{+110}_{-199}$ & $1.02^{+0.04}_{-0.03}$  & $547^{+41}_{-59}$ &     \\
1979B & K9 & $0.69^{+0.13}_{-0.23}$ &  $454^{+98}_{-154}$  & 1.00 &  $4398^{+340}_{-569}$ & $0.70^{+0.04}_{-0.06}$  & $565^{+119}_{-182}$ & 0.10\\
1989A & F9 & $1.14^{+0.04}_{-0.03}$ &  $525^{+49}_{-23}$   &      &  $5948^{+99}_{-124}$  & $1.02^{+0.03}_{-0.03}$  & $600^{+48}_{-40}$ &     \\
1989B & K2 & $0.90^{+0.03}_{-0.04}$ & $1299^{+195}_{-156}$ & 4.73 &  $5179^{+376}_{-447}$ & $0.86^{+0.14}_{-0.10}$  & $1536^{+353}_{-243}$ & 3.78\\
2001A & K1 & $0.97^{+0.01}_{-0.01}$ &  $266^{+8}_{-6}$     &      &  $5386^{+264}_{-271}$ & $0.93^{+0.09}_{-0.07}$  & $318^{+52}_{-40}$ &     \\
2001B & G5 & $0.93^{+0.05}_{-0.03}$ & $1761^{+278}_{-200}$ & 7.47 &  $5123^{+396}_{-351}$ & $0.90^{+0.13}_{-0.09}$  & $2092^{+498}_{-270}$ & 6.45\\
2009A & F7 & $1.18^{+0.03}_{-0.03}$ &  $696^{+59}_{-47}$   &      &  $6066^{+101}_{-155}$ & $1.05^{+0.03}_{-0.03}$  & $797^{+41}_{-82}$ &     \\
2009B & K4 & $0.82^{+0.07}_{-0.17}$ & $1204^{+196}_{-264}$ & 1.88 &  $4878^{+692}_{-381}$ & $0.82^{+0.33}_{-0.09}$  & $1715^{+767}_{-298}$ & 3.05\\
2059A & K2 & $0.93^{+0.02}_{-0.01}$ &  $246^{+18}_{-14}$   &      &  $5104^{+315}_{-202}$ & $0.91^{+0.11}_{-0.07}$  & $288^{+54}_{-25}$ &     \\
2059B & K5 & $0.85^{+0.02}_{-0.03}$ &  $268^{+22}_{-23}$   & 0.75 &  $4568^{+233}_{-288}$ & $0.73^{+0.06}_{-0.04}$  & $295^{+26}_{-37}$ & 0.11\\
2069A & F7 & $1.21^{+0.02}_{-0.03}$ &  $723^{+52}_{-54}$   &      &  $5947^{+103}_{-147}$ & $1.02^{+0.03}_{-0.03}$  & $713^{+62}_{-53}$ &     \\
2069B & K8 & $0.71^{+0.13}_{-0.26}$ & $1266^{+407}_{-397}$ & 1.36 &  $4912^{+404}_{-419}$ & $0.80^{+0.15}_{-0.10}$  & $1874^{+383}_{-275}$ & 4.12\\
2083A & G1 & $1.16^{+0.10}_{-0.04}$ &  $658^{+199}_{-57}$  &      &  $5889^{+152}_{-211}$ & $1.01^{+0.03}_{-0.03}$  & $685^{+66}_{-88}$ &     \\
2083B & F7 & $1.09^{+0.18}_{-0.38}$ & $1467^{+601}_{-923}$ & 0.86 &  $5139^{+1022}_{-481}$& $0.85^{+0.26}_{-0.10}$  & $1178^{+697}_{-319}$ & 1.51\\
2117A & K5 & $0.86^{+0.01}_{-0.01}$ &  $609^{+21}_{-26}$   &      &                       &                         &     \\
2117B & K3 & $0.88^{+0.01}_{-0.01}$ &  $818^{+31}_{-33}$   & 5.34 &  $4640^{+768}_{-333}$ & $0.75^{+0.48}_{-0.08}$  & $988^{+436}_{-178}$ &        \\
2143A & G2 & $1.07^{+0.03}_{-0.03}$ &  $643^{+39}_{-36}$   &      &  $5757^{+133}_{-206}$ & $1.00^{+0.02}_{-0.03}$  & $717^{+72}_{-74}$ &     \\
2143B & A0 & $1.19^{+0.21}_{-0.14}$ &$4408^{+4189}_{-1290}$& 2.92 &  $5778^{+257}_{-364}$ & $0.99^{+0.06}_{-0.08}$  & $3590^{+667}_{-559}$ & 5.10\\
2159A & F5 & $1.26^{+0.01}_{-0.02}$ &  $745^{+30}_{-45}$   &      &  $6130^{+87}_{-112}$  & $1.06^{+0.04}_{-0.03}$  & $739^{+30}_{-44}$ &     \\
2159B & M0 & $0.65^{+0.24}_{-0.27}$ &  $742^{+373}_{-326}$ & 0.01 &  $4398^{+340}_{-569}$ & $0.70^{+0.04}_{-0.06}$  & $961^{+193}_{-306}$ & 0.72\\
2247A & K3 & $0.90^{+0.01}_{-0.01}$ &  $337^{+11}_{-20}$   &      &  $4954^{+1280}_{-222}$& $0.79^{+0.48}_{-0.07}$  & $528^{+383}_{-143}$ &     \\
2247B & M0 & $0.53^{+0.11}_{-0.10}$ & $1116^{+234}_{-234}$ & 3.33 &                       &                         &        \\
2289A & F3 & $1.30^{+0.01}_{-0.01}$ &  $842^{+23}_{-27}$   &      &  $6201^{+64}_{-137}$  & $1.06^{+0.02}_{-0.02}$  & $753^{+32}_{-42}$ &     \\
2289B & K6 & $0.80^{+0.05}_{-0.08}$ & $1094^{+130}_{-140}$ & 1.78 &  $4620^{+807}_{-349}$ & $0.72^{+0.35}_{-0.08}$  & $1540^{+715}_{-296}$ & 2.64\\
2317A & F7 & $1.17^{+0.03}_{-0.05}$ &  $817^{+58}_{-64}$   &      &  $5969^{+106}_{-128}$ & $1.01^{+0.03}_{-0.03}$  & $904^{+64}_{-67}$ &     \\
2317B & G7 & $0.88^{+0.02}_{-0.03}$ & $2661^{+219}_{-195}$ & 9.06 &  $4650^{+459}_{-296}$ & $0.75^{+0.19}_{-0.06}$  & $3029^{+603}_{-392}$ & 5.35\\
2363A & K0 & $0.95^{+0.01}_{-0.01}$ &  $448^{+17}_{-19}$   &      &  $5386^{+360}_{-179}$ & $0.95^{+0.14}_{-0.07}$  & $578^{+129}_{-69}$ &     \\
2363B & K3 & $0.65^{+0.40}_{-0.30}$ &$3298^{+5208}_{-1596}$& 1.79 &  $4923^{+515}_{-384}$ & $0.83^{+0.21}_{-0.08}$  & $4826^{+1448}_{-708}$ & 5.90\\
2377A & F9 & $1.08^{+0.10}_{-0.09}$ &  $890^{+194}_{-181}$ &      &                       &                         &     \\
2377B & K1 & $0.92^{+0.05}_{-0.06}$ &  $831^{+141}_{-104}$ & 0.26 &                       &                         &        \\
2377C & K2 & $0.64^{+0.38}_{-0.34}$ &$2449^{+3109}_{-1566}$& 0.99 &                       &                         &        \\
2377D & K7 & $0.50^{+0.34}_{-0.22}$ & $1767^{+1465}_{-883}$& 0.97 &                       &                         &        \\
2413A & G7 & $0.98^{+0.04}_{-0.03}$ &  $756^{+78}_{-53}$   &      &  $5500^{+269}_{-258}$ & $0.99^{+0.08}_{-0.07}$  & $935^{+155}_{-127}$ &     \\
2413B & M2 & $0.48^{+0.21}_{-0.14}$ &  $324^{+136}_{-113}$ & 2.96 &                       &                         &        \\
2443A & F5 & $1.21^{+0.03}_{-0.04}$ &  $838^{+77}_{-101}$  &      &  $6141^{+87}_{-88}$   & $1.05^{+0.04}_{-0.03}$  & $934^{+33}_{-50}$ &     \\
2443B & K6 & $0.70^{+0.12}_{-0.22}$ & $1659^{+361}_{-504}$ & 1.61 &  $4742^{+878}_{-576}$ & $0.81^{+0.33}_{-0.12}$  & $2790^{+1614}_{-744}$ & 2.49\\
2542A & M0 & $0.65^{+0.04}_{-0.08}$ &  $278^{+18}_{-28}$   &      &                       &                         &     \\
2542B & M3 & $0.38^{+0.07}_{-0.06}$ &  $218^{+49}_{-45}$   & 1.06 &                       &                         &        \\
2601A & F3 & $1.28^{+0.03}_{-0.05}$ & $1133^{+103}_{-141}$ &      &  $6214^{+75}_{-124}$  & $1.06^{+0.03}_{-0.03}$  & $1086^{+46}_{-58}$ &     \\
2601B & G2 & $1.07^{+0.19}_{-0.11}$ & $1193^{+560}_{-299}$ & 0.19 &  $5708^{+286}_{-393}$ & $0.98^{+0.06}_{-0.09}$  & $1283^{+249}_{-234}$ & 0.83\\
2601C & G7 & $0.93^{+0.07}_{-0.04}$ & $2076^{+411}_{-212}$ & 4.00 &  $4846^{+406}_{-299}$ & $0.78^{+0.14}_{-0.09}$  & $2236^{+443}_{-250}$ & 4.52\\
2601D &    &                        &                      &      &  $5214^{+374}_{-568}$ & $0.90^{+0.08}_{-0.13}$  & $6068^{+1393}_{-1159}$ & 4.30\\
2657A & G0 & $1.09^{+0.07}_{-0.05}$ &  $504^{+61}_{-48}$   &      &  $5859^{+127}_{-213}$ & $1.00^{+0.02}_{-0.03}$  & $585^{+56}_{-67}$ &     \\
2657B & G6 & $1.01^{+0.05}_{-0.02}$ &  $454^{+54}_{-32}$   & 0.69 &  $5688^{+249}_{-269}$ & $1.00^{+0.06}_{-0.07}$  & $566^{+92}_{-78}$ & 0.17\\
2664A & K0 & $0.94^{+0.01}_{-0.01}$ &  $868^{+50}_{-43}$   &      &  $5134^{+359}_{-167}$ & $0.87^{+0.15}_{-0.06}$  & $1025^{+208}_{-101}$ &     \\
2664B & F6 & $1.01^{+0.03}_{-0.03}$ & $1708^{+198}_{-136}$ & 5.80 &  $5620^{+252}_{-255}$ & $0.98^{+0.06}_{-0.07}$  & $2112^{+340}_{-286}$ & 3.07\\
2681A & F7 & $1.04^{+0.04}_{-0.03}$ & $1469^{+160}_{-133}$ &      &  $5763^{+202}_{-287}$ & $0.98^{+0.06}_{-0.06}$  & $1774^{+251}_{-254}$ &     \\
2681B & K3 & $0.88^{+0.01}_{-0.02}$ & $1191^{+68}_{-62}$   & 1.86 &  $4650^{+523}_{-340}$ & $0.75^{+0.26}_{-0.06}$  & $1366^{+332}_{-210}$ & 0.98\\
2705A & M3 & $0.39^{+0.13}_{-0.07}$ &   $77^{+28}_{-16}$   &      &                       &                         &     \\
2705B & M5 & $0.30^{+0.08}_{-0.01}$ &  $186^{+36}_{-11}$   & 3.62 &                       &                         &        \\
2711A & F2 & $1.28^{+0.02}_{-0.03}$ & $1193^{+77}_{-93}$   &      &  $6059^{+61}_{-129}$  & $1.03^{+0.03}_{-0.03}$  & $1072^{+40}_{-94}$ &     \\
2711B & F2 & $1.27^{+0.02}_{-0.03}$ & $1241^{+94}_{-115}$  & 0.35 &  $6059^{+57}_{-132}$  & $1.03^{+0.03}_{-0.04}$  & $1131^{+42}_{-99}$ & 0.55\\
2722A & F3 & $1.29^{+0.01}_{-0.01}$ &  $808^{+21}_{-20}$   &      &  $6188^{+86}_{-158}$  & $1.05^{+0.02}_{-0.02}$  & $732^{+41}_{-47}$ &     \\
2722B & K8 & $0.69^{+0.06}_{-0.09}$ & $1341^{+132}_{-167}$ & 3.17 &                       &                         &        \\
2779A & F3 & $1.27^{+0.03}_{-0.04}$ & $1573^{+163}_{-161}$ &      &  $6140^{+74}_{-106}$  & $1.05^{+0.03}_{-0.04}$  & $1502^{+52}_{-86}$ &     \\
2779B & G8 & $0.95^{+0.12}_{-0.06}$ & $1793^{+608}_{-260}$ & 0.72 &  $5111^{+384}_{-373}$ & $0.87^{+0.12}_{-0.09}$  & $1991^{+472}_{-263}$ & 1.82\\
2813A & K3 & $0.93^{+0.01}_{-0.01}$ &  $310^{+14}_{-12}$   &      &  $5128^{+443}_{-224}$ & $0.88^{+0.20}_{-0.07}$  & $383^{+106}_{-47}$ &     \\
2813B & F8 & $1.25^{+0.05}_{-0.09}$ & $1466^{+215}_{-265}$ & 4.36 &  $6049^{+122}_{-159}$ & $1.03^{+0.04}_{-0.03}$  & $1363^{+81}_{-140}$ & 5.58\\
2813C & M1 & $0.57^{+0.23}_{-0.21}$ &$3541^{+1995}_{-1484}$& 2.18 &  $4887^{+1083}_{-835}$& $0.89^{+0.32}_{-0.22}$  & $7753^{+5561}_{-2760}$ & 2.67\\
2837A & F0 & $1.40^{+0.05}_{-0.04}$ & $1430^{+125}_{-105}$ &      &                       &                         &     \\
2837B & F0 & $1.38^{+0.07}_{-0.04}$ & $1527^{+192}_{-122}$ & 0.56 &                       &                         &        \\
2859A & G8 & $0.98^{+0.02}_{-0.01}$ &  $432^{+18}_{-9}$    &      &  $5582^{+196}_{-256}$ & $0.95^{+0.06}_{-0.07}$  & $546^{+81}_{-63}$ &     \\
2859B & A7 & $1.05^{+0.16}_{-0.22}$ & $2142^{+822}_{-896}$ & 1.91 &  $6169^{+140}_{-137}$ & $1.05^{+0.05}_{-0.03}$  & $2914^{+224}_{-180}$ & 12.00\\
2869A & F4 & $1.28^{+0.01}_{-0.01}$ &  $919^{+29}_{-36}$   &      &  $6154^{+58}_{-145}$  & $1.05^{+0.03}_{-0.03}$  & $840^{+31}_{-60}$ &     \\
2869B & K4 & $0.63^{+0.09}_{-0.10}$ & $3401^{+481}_{-559}$ & 4.43 &                       &                         &        \\
2904A & F5 & $1.27^{+0.01}_{-0.02}$ &  $587^{+27}_{-28}$   &      &  $6116^{+63}_{-212}$  & $1.05^{+0.03}_{-0.04}$  & $537^{+26}_{-68}$ &     \\
2904B & A1 & $1.28^{+0.11}_{-0.26}$ & $1976^{+470}_{-552}$ & 2.51 &                       &                         &        \\
2971A & F4 & $1.29^{+0.01}_{-0.01}$ &  $622^{+18}_{-22}$   &      &  $6188^{+86}_{-158}$  & $1.05^{+0.02}_{-0.02}$  & $569^{+32}_{-37}$ &     \\
2971B & G2 & $0.96^{+0.04}_{-0.06}$ & $1712^{+236}_{-331}$ & 3.29 &  $5720^{+216}_{-289}$ & $0.99^{+0.05}_{-0.07}$  & $2281^{+363}_{-310}$ & 5.49\\
2971C & K7 & $0.47^{+0.33}_{-0.34}$ &$2446^{+2884}_{-1060}$& 1.72 &  $5605^{+373}_{-590}$ & $0.96^{+0.08}_{-0.14}$  & $6324^{+1639}_{-1411}$ & 4.08\\
3020A & F3 & $1.31^{+0.01}_{-0.01}$ &  $956^{+31}_{-28}$   &      &                       &                         &     \\
3020B & K9 & $0.77^{+0.03}_{-0.05}$ &  $526^{+34}_{-37}$   & 9.76 &                       &                         &        \\
3020C & K3 & $0.77^{+0.07}_{-0.14}$ & $3032^{+479}_{-508}$ & 4.08 &                       &                         &        \\
3069A & F4 & $1.20^{+0.04}_{-0.05}$ & $1227^{+162}_{-188}$ &      &  $6110^{+89}_{-83}$   & $1.04^{+0.04}_{-0.03}$  & $1402^{+51}_{-62}$ &     \\
3069B & K2 & $0.90^{+0.01}_{-0.02}$ & $1164^{+58}_{-60}$   & 0.32 &  $4876^{+173}_{-134}$ & $0.81^{+0.04}_{-0.04}$  & $1332^{+87}_{-55}$ & 0.66\\
3106A & G5 & $1.04^{+0.05}_{-0.03}$ & $1189^{+142}_{-110}$ &      &  $5855^{+202}_{-287}$ & $1.00^{+0.05}_{-0.06}$  & $1526^{+207}_{-228}$ &     \\
3106B & A3 & $1.29^{+0.05}_{-0.06}$ & $3533^{+582}_{-576}$ & 3.95 &  $6188^{+88}_{-109}$  & $1.05^{+0.04}_{-0.03}$  & $3253^{+155}_{-151}$ & 6.74\\
3377A & G9 & $0.94^{+0.01}_{-0.01}$ &  $675^{+24}_{-38}$   &      &  $5110^{+350}_{-160}$ & $0.94^{+0.16}_{-0.05}$  & $784^{+158}_{-73}$ &     \\
3377B & M7 & $0.29^{+0.01}_{-0.01}$ &  $271^{+15}_{-14}$   & 9.89 &                       &                         &        \\
3377C & M2 & $0.36^{+0.09}_{-0.07}$ & $1215^{+405}_{-296}$ & 1.82 &                       &                         &        \\
3401A & G8 & $1.01^{+0.03}_{-0.02}$ &  $665^{+60}_{-41}$   &      &  $5526^{+252}_{-277}$ & $0.93^{+0.07}_{-0.07}$  & $769^{+122}_{-110}$ &     \\
3401B & B8 & $\geq1.40            $ & $\geq3510$           & 6.69 &                       &                         &        \\
4004A & F8 & $1.14^{+0.04}_{-0.04}$ &  $400^{+32}_{-24}$   &      &  $5972^{+102}_{-135}$ & $1.02^{+0.02}_{-0.02}$  & $456^{+32}_{-35}$ &     \\
4004B & K8 & $0.75^{+0.05}_{-0.05}$ &  $524^{+48}_{-43}$   & 2.31 &                       &                         &        \\
4209A & F3 & $1.09^{+0.11}_{-0.07}$ & $1647^{+385}_{-264}$ &      &  $6021^{+119}_{-135}$ & $1.02^{+0.03}_{-0.03}$  & $2124^{+125}_{-191}$ &     \\
4209B & K7 & $0.83^{+0.22}_{-0.43}$ & $1859^{+6446}_{-929}$& 0.21 &  $4887^{+1083}_{-835}$& $0.89^{+0.32}_{-0.22}$  & $1814^{+1305}_{-654}$ & 0.24\\
4292A & G4 & $1.06^{+0.03}_{-0.02}$ &  $360^{+17}_{-16}$   &      &  $5857^{+140}_{-213}$ & $1.00^{+0.03}_{-0.03}$  & $436^{+43}_{-52}$ &     \\
4292B & M6 & $0.29^{+0.05}_{-0.01}$ &  $610^{+48}_{-38}$   & 6.01 &                       &                         &        \\
4331A & F2 & $1.34^{+0.20}_{-0.13}$ & $1177^{+499}_{-283}$ &      &  $6140^{+140}_{-119}$ & $1.04^{+0.04}_{-0.03}$  & $964^{+76}_{-51}$ &     \\
4331B & F3 & $1.28^{+0.15}_{-0.15}$ & $1109^{+404}_{-299}$ & 0.14 &  $6033^{+136}_{-134}$ & $1.03^{+0.03}_{-0.03}$  & $962^{+58}_{-90}$ & 0.03\\
4407A & F8 & $1.17^{+0.05}_{-0.06}$ &  $242^{+29}_{-24}$   &      &                       &                         &     \\
4407A & G2 & $1.14^{+0.05}_{-0.05}$ &  $230^{+29}_{-18}$   & 0.32 &                       &                         &        \\
4407B & K5 & $0.82^{+0.06}_{-0.28}$ &  $278^{+36}_{-87}$   & 0.39 &                       &                         &        \\
4407C &    &                        &                      &      &                       &                         &        \\
4463A & K8 & $0.79^{+0.01}_{-0.02}$ &  $427^{+14}_{-14}$   &      &                       &                         &     \\
4463B & K5 & $0.85^{+0.02}_{-0.02}$ &  $543^{+25}_{-28}$   & 3.71 &                       &                         &        \\
4634A & A8 & $1.33^{+0.05}_{-0.02}$ & $1223^{+136}_{-74}$  &      &                       &                         &     \\
4634B & K2 & $0.92^{+0.09}_{-0.06}$ &  $669^{+202}_{-84}$  & 2.58 &  $4670^{+403}_{-349}$ & $0.73^{+0.13}_{-0.09}$  & $668^{+116}_{-90}$ &        \\
4768A & G4 & $1.01^{+0.02}_{-0.02}$ & $1037^{+91}_{-77}$   &      &  $5696^{+250}_{-282}$ & $0.99^{+0.07}_{-0.08}$  & $1330^{+237}_{-182}$ &     \\
4768B & K5 & $0.73^{+0.09}_{-0.15}$ & $2014^{+578}_{-507}$ & 1.90 &                       &                         &        \\
4822A & F6 & $1.27^{+0.05}_{-0.03}$ &  $769^{+109}_{-67}$  &      &  $6127^{+83}_{-84}$   & $1.03^{+0.03}_{-0.04}$  & $733^{+25}_{-35}$ &     \\
4822B & K9 & $0.60^{+0.18}_{-0.17}$ & $1665^{+537}_{-564}$ & 1.56 &                       &                         &        \\
4871A & F4 & $1.28^{+0.01}_{-0.01}$ &  $724^{+18}_{-25}$   &      &                       &                         &     \\
4871B & A5 & $1.22^{+0.08}_{-0.11}$ & $2540^{+545}_{-546}$ & 3.32 &  $6030^{+133}_{-134}$ & $1.03^{+0.03}_{-0.03}$  & $2556^{+155}_{-236}$ &        \\
5578A & G5 & $1.11^{+0.05}_{-0.04}$ &  $190^{+17}_{-10}$   &      &  $6005^{+114}_{-157}$ & $1.00^{+0.02}_{-0.03}$  & $233^{+16}_{-21}$ &     \\
5578B & G5 & $1.07^{+0.12}_{-0.07}$ &  $383^{+99}_{-61}$   & 3.05 &  $5778^{+201}_{-296}$ & $0.99^{+0.05}_{-0.06}$  & $438^{+70}_{-58}$ & 3.41\\
5762A & G6 & $0.99^{+0.03}_{-0.03}$ & $1130^{+146}_{-86}$  &      &  $5551^{+255}_{-293}$ & $0.95^{+0.07}_{-0.08}$  & $1385^{+230}_{-199}$ &     \\
5762B & F4 & $1.08^{+0.12}_{-0.07}$ & $2042^{+556}_{-342}$ & 2.45 &  $5886^{+195}_{-303}$ & $1.02^{+0.05}_{-0.06}$  & $2431^{+314}_{-400}$ & 2.27\\

\end{longtable*}

\subsection{On Potential Giants}
Dwarf-giant eclipsing binaries were originally expected to be approximately 200 times more abundant than detected planet-star transits in the \textit{Kepler} field \citep{brown2003}. The assembly of the \textit{Kepler} Input Catalog made use of Bayesian techniques to exclude most of these giants \citep{brown2011}. Remaining dwarf-giant eclipsing binaries are identified and screened by the \textit{Kepler} analysis pipeline, particularly by the presence of secondary eclipses in light curves \citep{batalha2010, tenenbaum2013}. Although a dwarf-giant eclipsing binary with (or as) a contaminating companion should be a relatively rare arrangement, it is reasonable to expect that some might still fall out of a dataset as large as the KIC. It has been demonstrated spectroscopically that a number of late-type \textit{Kepler} Input Catalog stars photometrically characterized as dwarfs are in fact giants, but also that an improved photometric cut exists as $K_p - J > 2$ and $K_p < 14$ that contains $96\% \pm 1\%$ giants (with the corresponding $K_p - J > 2$ and $K_p > 14$ containing only $7\% \pm 3\%$), allowing us to investigate our updated photometry for the possibility of giants hiding in blended KOIs \citep{mann2012}. None of our observed stars meet this set of criteria, and thus all objects are likely dwarfs.

\subsection{Probability of Physical Association}\label{multiplicity}

Table~\ref{stellarparams} lists the confidence for each host/companion pair to be physically unassociated, derived from the respective distances and uncertainties of both fitting methods. The noted uncertainties in photometrically fitting temperature/radius ($\pm200$K and 0.2dex, respectively) to individual stars reported in B11 are ignored for the distance estimates and physical association confidences as they are functions of stellar age and metallicity, which should be consistent across all members of a physically associated system. We treat all pairs with $\geq5\sigma$ level of confidence to be inconsistent with a physically associated/gravitationally bound scenario. Note that pairs with $\leq5\sigma$ are not necessarily associated/bound but are not inconsistent with such an interpretation. The KH07 method then identifies 13 physically unassociated companions, while the B11 method finds 10.

The two methods agree to $5\sigma$ on the unboundedness of only one companion, 2001B. Of the other 12 unbound candidates via KH07, 9 do not have B11 fits, and one (2317B) has a marginal B11 $\sigma_{unassoc} = 4.90$. Only two, 1989B and 2664B, are unbound by KH07 and disputed by B11, for which the methods agree on distances but have respective B11 $\sigma_{unassoc}$ of 3.63 and 2.91 due to larger uncertainties on the B11 results.

B11 also identifies 10 additional physically unassociated companions that do not qualify by the KH07 measurement, though two (628C and 2813B) are marginal. We note that uncertainties for the B11 method are systematically underestimated by the granular dataset, and many could not be fit due to that catalog's relative sparsity, particularly for late-type stars.

The B11 results are presented to check the reproducibility of the KH07 fitting method, but given the noted issues with the former the KH07 results are preferred, and are the focus of this work.

\begin{longtable*}{ r | c c | c c | c c | c c }
\caption{Adjusted Transit Depth and Candidate Sizes}\label{planets}\\
\setlength{\extrarowheight}{3pt}
 & \multicolumn{2}{c|}{A} & \multicolumn{2}{c|}{B} & \multicolumn{2}{c|}{C} & \multicolumn{2}{c}{D}\\
	object & depth (mmag) & $R_{\earth}$ & depth(mmag) & $R_{\earth}$ &	depth(mmag) & $R_{\earth}$ & depth(mmag) & $R_{\earth}$ \\
	\hline
	\endfirsthead
 & \multicolumn{2}{c|}{A} & \multicolumn{2}{c|}{B} & \multicolumn{2}{c|}{C} & \multicolumn{2}{c}{D}\\
	object & depth (mmag) & $R_{\earth}$ & depth(mmag) & $R_{\earth}$ &	depth(mmag) & $R_{\earth}$ & depth(mmag) & $R_{\earth}$ \\
	\hline
	\endhead
	\endfoot
	\endlastfoot
  0190.01  &  16.61  $\pm$  0.026  & $15.99^{+0.54}_{-0.94}$ &  57.24  $\pm$  0.092  & $22.98^{+0.99}_{-0.49}$ \\
  0191.01  &  18.58  $\pm$  0.027  & $15.2^{+0.71}_{-0.43}$ &  345.0  $\pm$  0.502  & $60.31^{+7.41}_{-4.56}$ \\
  0191.02  &  0.840  $\pm$  0.013  & $3.24^{+0.15}_{-0.09}$ &  13.55  $\pm$  0.223  & $12.87^{+1.6}_{-0.99}$ \\
  0191.03  &  0.232  $\pm$  0.009  & $1.71^{+0.08}_{-0.05}$ &  3.739  $\pm$  0.147  & $6.78^{+0.86}_{-0.53}$ \\
  0191.04  &  0.725  $\pm$  0.030  & $3.01^{+0.15}_{-0.09}$ &  11.69  $\pm$  0.494  & $11.96^{+1.53}_{-0.94}$ \\
  0268.01  &  0.546  $\pm$  0.003  & $3.13^{+0.1}_{-0.12}$ &  127.9  $\pm$  0.918  & $30.54^{+1.46}_{-2.56}$ &  N/A  & $\geq 100.3$ \\
  0401.01  &  2.449  $\pm$  0.016  & $5.44^{+0.16}_{-0.1}$ &  35.30  $\pm$  0.238  & $15.21^{+0.98}_{-1.77}$ \\
  0401.02  &  1.868  $\pm$  0.043  & $4.75^{+0.14}_{-0.09}$ &  26.83  $\pm$  0.618  & $13.29^{+0.87}_{-1.57}$ \\
  0401.03  &  0.405  $\pm$  0.019  & $2.21^{+0.07}_{-0.04}$ &  5.775  $\pm$  0.282  & $6.19^{+0.42}_{-0.75}$ \\
  0425.01  &  23.09  $\pm$  0.065  & $19.46^{+0.79}_{-0.63}$ &  51.76  $\pm$  0.147  & $28.23^{+1.42}_{-1.65}$ \\
  0511.01  &  0.757  $\pm$  0.009  & $3.6^{+0.09}_{-0.15}$ &  13.15  $\pm$  0.173  & $10.05^{+0.48}_{-0.97}$ &  278.9  $\pm$  3.670  & $30.62^{+7.36}_{-7.88}$ \\
  0511.02  &  0.210  $\pm$  0.008  & $1.9^{+0.05}_{-0.08}$ &  3.645  $\pm$  0.146  & $5.3^{+0.26}_{-0.52}$ &  70.73  $\pm$  2.846  & $16.16^{+3.98}_{-4.26}$ \\
  0688.01  &  0.354  $\pm$  0.006  & $2.56^{+0.02}_{-0.04}$ &  2.608  $\pm$  0.048  & $5.39^{+0.43}_{-0.27}$ \\
  0984.01  &  2.187  $\pm$  0.013  & $5.04^{+0.15}_{-0.1}$ &  2.376  $\pm$  0.014  & $5.25^{+0.15}_{-0.1}$ \\
  0987.01  &  0.232  $\pm$  0.005  & $1.61^{+0.02}_{-0.02}$ &  6.369  $\pm$  0.139  & $6.42^{+0.26}_{-0.43}$ \\
  1066.01  &  15.39  $\pm$  0.032  & $14.62^{+0.65}_{-0.65}$ &  1261.  $\pm$  2.662  & $76.84^{+2.72}_{-3.62}$ \\
  1067.01  &  50.95  $\pm$  0.069  & $30.12^{+0.23}_{-0.23}$ &          N/A          & $\geq93.78$ \\
  1112.01  &  0.689  $\pm$  0.022  & $3.38^{+0.11}_{-0.17}$ &  50.69  $\pm$  1.648  & $17.0^{+1.2}_{-0.96}$ \\
  1214.01  &  0.294  $\pm$  0.018  & $1.79^{+0.04}_{-0.04}$ &  0.890  $\pm$  0.056  & $5.0^{+0.3}_{-0.33}$ \\
  1447.01  &  228.6  $\pm$  0.074  & $68.91^{+1.9}_{-1.43}$ &          N/A          & $\geq97.05$ \\
  1447.02  &  17.96  $\pm$  0.038  & $20.25^{+0.56}_{-0.42}$ &  125.5  $\pm$  0.267  & $32.79^{+1.08}_{-0.72}$ \\
  1700.01  &  0.442  $\pm$  0.016  & $2.13^{+0.05}_{-0.05}$ &  1.174  $\pm$  0.042  & $3.19^{+0.11}_{-0.11}$ \\
  1784.01  &  7.479  $\pm$  0.092  & $10.48^{+0.46}_{-0.55}$ &  12.74  $\pm$  0.157  & $14.96^{+0.36}_{-0.6}$ \\
  1880.01  &  0.680  $\pm$  0.009  & $2.13^{+0.03}_{-0.14}$ &  18.33  $\pm$  0.248  & $13.83^{+1.29}_{-0.71}$ \\
  1884.01  &  3.201  $\pm$  0.049  & $6.27^{+0.42}_{-0.24}$ &  123.3  $\pm$  1.919  & $26.09^{+5.08}_{-9.79}$ &  228.7  $\pm$  3.558  & $27.57^{+11.09}_{-9.64}$ &  356.4  $\pm$  5.543  & $23.07^{+10.38}_{-5.77}$ \\
  1884.02  &  0.618  $\pm$  0.039  & $2.76^{+0.19}_{-0.11}$ &  22.80  $\pm$  1.459  & $11.47^{+2.33}_{-4.5}$ &  40.66  $\pm$  2.603  & $12.13^{+5.1}_{-4.44}$ &  60.42  $\pm$  3.868  & $10.15^{+4.84}_{-2.69}$ \\
  1916.01  &  0.395  $\pm$  0.009  & $2.73^{+0.02}_{-0.02}$ &  4.866  $\pm$  0.115  & $7.0^{+0.45}_{-0.37}$ \\
  1916.02  &  0.305  $\pm$  0.006  & $2.39^{+0.02}_{-0.02}$ &  3.754  $\pm$  0.079  & $6.15^{+0.39}_{-0.33}$ \\
  1916.03  &  0.079  $\pm$  0.004  & $1.22^{+0.01}_{-0.01}$ &  0.980  $\pm$  0.051  & $3.14^{+0.21}_{-0.17}$ \\
  1989.01  &  0.534  $\pm$  0.020  & $2.76^{+0.1}_{-0.08}$ &  13.31  $\pm$  0.505  & $11.08^{+0.37}_{-0.37}$ \\
  2001.01  &  0.205  $\pm$  0.007  & $1.45^{+0.02}_{-0.02}$ &  13.88  $\pm$  0.502  & $11.43^{+0.64}_{-0.38}$ \\
  2009.01  &  0.626  $\pm$  0.022  & $3.06^{+0.08}_{-0.11}$ &  25.22  $\pm$  0.901  & $13.72^{+1.2}_{-3.59}$ \\
  2059.01  &  0.186  $\pm$  0.007  & $1.33^{+0.03}_{-0.01}$ &  0.509  $\pm$  0.020  & $2.01^{+0.05}_{-0.07}$ \\
  2059.02  &  0.057  $\pm$  0.005  & $0.74^{+0.02}_{-0.01}$ &  0.156  $\pm$  0.015  & $1.11^{+0.03}_{-0.04}$ \\
  2069.01  &  0.678  $\pm$  0.013  & $3.3^{+0.06}_{-0.08}$ &  29.60  $\pm$  0.605  & $12.7^{+2.19}_{-4.92}$ \\
  2083.01  &  0.399  $\pm$  0.015  & $2.49^{+0.17}_{-0.13}$ &  1.027  $\pm$  0.039  & $4.16^{+0.24}_{-1.01}$ \\
  2117.01  &  1.519  $\pm$  0.074  & $3.51^{+0.04}_{-0.04}$ &  2.060  $\pm$  0.101  & $4.18^{+0.05}_{-0.05}$ \\
  2247.01  &  0.205  $\pm$  0.011  & $1.35^{+0.02}_{-0.02}$ &  20.34  $\pm$  1.121  & $8.02^{+1.72}_{-1.72}$ \\
  2289.01  &  0.369  $\pm$  0.018  & $2.61^{+0.02}_{-0.02}$ &  28.81  $\pm$  1.470  & $14.12^{+0.92}_{-1.48}$ \\
  2289.02  &  0.175  $\pm$  0.009  & $1.8^{+0.01}_{-0.01}$ &  13.57  $\pm$  0.761  & $9.72^{+0.64}_{-1.02}$ \\
  2317.01  &  0.149  $\pm$  0.009  & $1.5^{+0.04}_{-0.07}$ &  16.21  $\pm$  1.076  & $11.68^{+0.28}_{-0.42}$ \\
  2363.01  &  0.204  $\pm$  0.012  & $1.42^{+0.02}_{-0.02}$ &  45.44  $\pm$  2.821  & $14.13^{+8.65}_{-6.78}$ \\
  2413.01  &  0.531  $\pm$  0.029  & $2.39^{+0.15}_{-0.08}$ &  3.329  $\pm$  0.184  & $2.53^{+1.27}_{-0.89}$ \\
  2413.02  &  0.457  $\pm$  0.038  & $2.22^{+0.14}_{-0.07}$ &  2.868  $\pm$  0.238  & $2.35^{+1.21}_{-0.85}$ \\
  2443.01  &  0.110  $\pm$  0.007  & $1.33^{+0.04}_{-0.05}$ &  13.41  $\pm$  0.877  & $8.7^{+1.41}_{-2.82}$ \\
  2443.02  &  0.105  $\pm$  0.008  & $1.3^{+0.03}_{-0.05}$ &  12.79  $\pm$  1.084  & $8.5^{+1.4}_{-2.81}$ \\
  2542.01  &  0.576  $\pm$  0.033  & $1.63^{+0.13}_{-0.21}$ &  1.710  $\pm$  0.100  & $1.64^{+0.37}_{-0.27}$ \\
  2657.01  &  0.091  $\pm$  0.007  & $1.09^{+0.08}_{-0.05}$ &  0.117  $\pm$  0.010  & $1.16^{+0.07}_{-0.04}$ \\
  2664.01  &  1.377  $\pm$  0.105  & $3.65^{+0.04}_{-0.04}$ &  2.954  $\pm$  0.226  & $5.74^{+0.18}_{-0.12}$ \\
  2681.01  &  8.139  $\pm$  0.129  & $9.8^{+0.38}_{-0.29}$ &  26.00  $\pm$  0.413  & $14.76^{+0.17}_{-0.34}$ \\
  2681.02  &  1.006  $\pm$  0.111  & $3.45^{+0.15}_{-0.11}$ &  3.193  $\pm$  0.353  & $5.2^{+0.07}_{-0.13}$ \\
  2705.01  &  0.745  $\pm$  0.027  & $1.31^{+0.53}_{-0.27}$ &  11.40  $\pm$  0.416  & $3.34^{+0.69}_{-0.12}$ \\
  2711.01  &  0.442  $\pm$  0.013  & $2.82^{+0.05}_{-0.07}$ &  0.493  $\pm$  0.015  & $2.95^{+0.05}_{-0.1}$ \\
  2711.02  &  0.351  $\pm$  0.016  & $2.51^{+0.04}_{-0.06}$ &  0.392  $\pm$  0.018  & $2.63^{+0.04}_{-0.09}$ \\
  2722.01  &  0.161  $\pm$  0.004  & $1.72^{+0.01}_{-0.01}$ &  60.00  $\pm$  1.601  & $17.45^{+1.55}_{-2.33}$ \\
  2722.02  &  0.154  $\pm$  0.005  & $1.68^{+0.01}_{-0.01}$ &  57.25  $\pm$  1.971  & $17.05^{+1.53}_{-2.3}$ \\
  2722.03  &  0.109  $\pm$  0.003  & $1.41^{+0.01}_{-0.01}$ &  40.34  $\pm$  1.439  & $14.37^{+1.29}_{-1.94}$ \\
  2722.04  &  0.115  $\pm$  0.005  & $1.45^{+0.01}_{-0.01}$ &  42.41  $\pm$  1.847  & $14.73^{+1.33}_{-2.0}$ \\
  2722.05  &  0.105  $\pm$  0.006  & $1.39^{+0.01}_{-0.01}$ &  38.80  $\pm$  2.471  & $14.1^{+1.3}_{-1.95}$ \\
  2779.01  &  0.592  $\pm$  0.028  & $3.23^{+0.08}_{-0.08}$ &  6.113  $\pm$  0.292  & $7.76^{+1.02}_{-0.51}$ \\
  2813.01  &  0.205  $\pm$  0.017  & $1.4^{+0.02}_{-0.02}$ &  0.506  $\pm$  0.042  & $2.97^{+0.13}_{-0.23}$ &  38.89  $\pm$  3.261  & $11.66^{+4.86}_{-4.64}$ \\
  2837.01  &  0.259  $\pm$  0.010  & $2.36^{+0.09}_{-0.07}$ &  0.320  $\pm$  0.013  & $2.59^{+0.14}_{-0.08}$ \\
  2849.01  &  0.205  $\pm$  0.014  & $1.38^{+0.02}_{-0.03}$ &  0.435  $\pm$  0.029  & $4.54^{+0.65}_{-0.81}$ \\
  2859.01  &  0.100  $\pm$  0.007  & $1.04^{+0.02}_{-0.01}$ &  0.707  $\pm$  0.054  & $2.53^{+0.42}_{-0.54}$ \\
  2859.02  &  0.068  $\pm$  0.006  & $0.86^{+0.02}_{-0.01}$ &  0.482  $\pm$  0.044  & $2.09^{+0.35}_{-0.45}$ \\
  2859.03  &  0.078  $\pm$  0.007  & $0.92^{+0.02}_{-0.01}$ &  0.556  $\pm$  0.053  & $2.25^{+0.38}_{-0.48}$ \\
  2859.04  &  0.077  $\pm$  0.006  & $0.91^{+0.02}_{-0.01}$ &  0.544  $\pm$  0.044  & $2.22^{+0.37}_{-0.47}$ \\
  2859.05  &  0.107  $\pm$  0.008  & $1.07^{+0.02}_{-0.01}$ &  0.756  $\pm$  0.062  & $2.62^{+0.43}_{-0.56}$ \\
  2869.01  &  0.139  $\pm$  0.009  & $1.58^{+0.01}_{-0.01}$ &          N/A          & $\geq58.89$ \\
  2904.01  &  0.144  $\pm$  0.005  & $1.6^{+0.01}_{-0.03}$ &  0.895  $\pm$  0.033  & $3.6^{+0.45}_{-0.71}$ \\
  2971.01  &  0.071  $\pm$  0.004  & $1.14^{+0.01}_{-0.01}$ &  2.705  $\pm$  0.155  & $5.28^{+0.23}_{-0.34}$ &  21.15  $\pm$  1.214  & $4.85^{+12.94}_{-0.8}$ \\
  2971.02  &  0.103  $\pm$  0.006  & $1.37^{+0.01}_{-0.01}$ &  3.938  $\pm$  0.260  & $6.36^{+0.28}_{-0.42}$ &  30.91  $\pm$  2.042  & $5.85^{+15.73}_{-0.97}$ \\
  3020.01  &  0.106  $\pm$  0.006  & $1.42^{+0.01}_{-0.01}$ &  2.122  $\pm$  0.134  & $3.71^{+0.15}_{-0.26}$ &  137.1  $\pm$  8.661  & $28.92^{+2.63}_{-4.88}$ \\
  3069.01  &  0.387  $\pm$  0.031  & $2.47^{+0.09}_{-0.13}$ &  2.996  $\pm$  0.244  & $5.15^{+0.06}_{-0.12}$ \\
  3377.01  &  0.476  $\pm$  0.044  & $2.15^{+0.02}_{-0.02}$ &  15.94  $\pm$  1.488  & $3.82^{+0.01}_{-0.01}$ &  100.9  $\pm$  9.425  & $11.37^{+2.92}_{-2.27}$ \\
  3401.01  &  0.200  $\pm$  0.022  & $1.5^{+0.05}_{-0.03}$ &  0.452  $\pm$  0.049  & $3.32^{+0.25}_{-0.22}$ \\
  3401.02  &  0.603  $\pm$  0.061  & $2.6^{+0.08}_{-0.06}$ &  1.362  $\pm$  0.139  & $5.75^{+0.42}_{-0.38}$ \\
  4004.01  &  0.151  $\pm$  0.010  & $1.47^{+0.05}_{-0.05}$ &  6.263  $\pm$  0.440  & $6.2^{+0.44}_{-0.44}$ \\
  4209.01  &  1.439  $\pm$  0.350  & $4.25^{+0.45}_{-0.4}$ &  13.51  $\pm$  3.292  & $13.95^{+2.43}_{-2.28}$ \\
  4292.01  &  0.045  $\pm$  0.004  & $0.75^{+0.03}_{-0.02}$ &  50.35  $\pm$  4.668  & $6.73^{+1.52}_{-0.0}$ \\
  4331.01  &  0.125  $\pm$  0.011  & $1.59^{+0.37}_{-0.19}$ &  0.157  $\pm$  0.013  & $1.72^{+0.38}_{-0.21}$ \\
  4463.01  &  0.372  $\pm$  0.024  & $1.6^{+0.02}_{-0.06}$ &  0.376  $\pm$  0.024  & $1.7^{+0.04}_{-0.04}$ \\
  4634.01  &  0.118  $\pm$  0.012  & $1.51^{+0.06}_{-0.03}$ &  0.618  $\pm$  0.066  & $2.39^{+0.29}_{-0.2}$ \\
  4768.01  &  0.598  $\pm$  0.062  & $2.58^{+0.06}_{-0.06}$ &  24.00  $\pm$  2.489  & $11.77^{+1.77}_{-2.66}$ \\
  4822.01  &  0.035  $\pm$  0.003  & $0.79^{+0.03}_{-0.02}$ &  19.46  $\pm$  2.083  & $8.57^{+2.89}_{-2.73}$ \\
  4871.01  &  0.029  $\pm$  0.004  & $0.73^{+0.01}_{-0.01}$ &  0.524  $\pm$  0.070  & $2.87^{+0.22}_{-0.3}$ \\
  4871.02  &  0.038  $\pm$  0.004  & $0.83^{+0.01}_{-0.01}$ &  0.671  $\pm$  0.081  & $3.25^{+0.24}_{-0.33}$ \\
  5578.01  &  0.193  $\pm$  0.025  & $1.62^{+0.08}_{-0.07}$ &  0.997  $\pm$  0.133  & $3.54^{+0.49}_{-0.26}$ \\
  5762.01  &  0.484  $\pm$  0.065  & $2.28^{+0.08}_{-0.08}$ &  0.876  $\pm$  0.118  & $3.35^{+0.42}_{-0.25}$ \\

\caption*{The transit depth and radius relative to potential host for all transit candidates, evaluated for association with all possible host stars. Evaluated only for KOIs with Kepler-band contrast observations. Candidates with radii lower limits indicate the depth of the eclipse is equal to or greater than the star's full light.}\\
\end{longtable*}

\begin{table}
\caption{Probability of $R > 15R_{\earth}$ for Each Planet Candidate}\label{FPprobs}
\centering
\begin{tabular}{ c c c c c c }
KOI & $P_{A}$ & $P_{B}$& $P_{C}$& $P_{D}$& $P_{total}$\\
\hline
0190.01$^{a}$ & 0.85  & 1.00 &       &       & 0.93 \\
0191.01       & 0.68  & 1.00 &       &       & 0.84 \\
0191.02       &   0   & 0.09 &       &       & 0.05 \\
0191.04       &   0   & 0.02 &       &       & 0.01 \\
0268.01       &   0   & 1.00 & 1.00  &       & 0.67 \\
0401.01$^{b}$ &   0   & 0.55 &       &       & 0.27 \\
0401.02$^{b}$ &   0   & 0.03 &       &       & 0.01 \\
0425.01       & 1.00  & 1.00 &       &       & 1.00 \\
0511.01$^{b}$ &   0   &   0  & 0.98  &       & 0.33 \\
0511.02$^{b}$ &   0   &   0  & 0.61  &       & 0.20 \\
1066.01       & 0.30  & 1.00 &       &       & 0.65 \\
1067.01       & 1.00  & 1.00 &       &       & 1.00 \\
1112.01$^{c}$ &   0   & 0.95 &       &       & 0.48 \\
1447.01$^{c}$ & 1.00  & 1.00 &       &       & 1.00 \\
1447.02       & 1.00  & 1.00 &       &       & 1.00 \\
1784.01       &   0   & 0.46 &       &       & 0.23 \\
1880.01       &   0   & 0.18 &       &       & 0.09 \\
1884.01       &   0   & 0.87 & 0.90  & 0.92  & 0.65 \\
1884.02       &   0   & 0.07 & 0.29  & 0.14  & 0.21 \\
2009.01       &   0   & 0.14 &       &       & 0.07 \\
2069.01       &   0   & 0.15 &       &       & 0.07 \\
2289.01$^{b}$ &   0   & 0.17 &       &       & 0.09 \\
2363.01       &   0   & 0.46 &       &       & 0.23 \\
2681.01				&   0   & 0.08 &       &       & 0.04 \\
2722.01$^{b}$ &   0   & 0.85 &       &       & 0.43 \\
2722.02$^{b}$ &   0   & 0.81 &       &       & 0.41 \\
2722.03$^{b}$ &   0   & 0.31 &       &       & 0.16 \\
2722.04$^{b}$ &   0   & 0.42 &       &       & 0.21 \\
2722.05       &   0   & 0.24 &       &       & 0.12 \\
2813.01       &   0   &   0  & 0.25  &       & 0.08 \\
2869.01       &   0   & 1.00 &       &       & 0.50 \\
2971.01       &   0   &   0  & 0.22  &       & 0.07 \\
2971.02       &   0   &   0  & 0.28  &       & 0.09 \\
3020.01       &   0   &   0  & 1.00  &       & 0.33 \\
3377.01       &   0   &   0  & 0.11  &       & 0.04 \\
4209.01       &   0   & 0.33 &       &       & 0.16 \\
4768.01       &   0   & 0.03 &       &       & 0.02 \\
4822.01       &   0   & 0.01 &       &       & 0.01 \\
\end{tabular}

$ ^{a}$ Disposition is FALSE POSITIVE in the Exoplanet Archive as of 18 Sep 2015.

$ ^{b}$ Disposition is CONFIRMED in the Exoplanet Archive as of 18 Sep 2015.

$ ^{c}$ Other literature indicates candidate is false positive.

Estimated probabilities that each KOI planet candidate has a radius $R > 15R_\Earth$, for each potential host and summed across all. Only candidates for with $P(R > 15R_\Earth) \geq 0.01$ are listed. For full transit depth and planet size estimates, see Table~\ref{planets}.
\end{table}

\subsection{Updated Transiting Object Parameters}
For KOI systems observed in the \textit{Kepler} band, we reinterpret the relative depth and size of all transit candidates in Table~\ref{planets}, relying on the KH07 results as new stellar characteristics. As we lack the ability to determine whether the primary or a companion is the host of the transiting object, all possible scenarios are presented. Note that these derivations require knowledge of each KOI component's luminosity in the transit band, and thus only candidates with resolved LP600 photometry are shown. As mentioned in section~\ref{observations}, the LP600 combined with the EMCCD's sensitivity curve approximates the \textit{Kepler} passband, suppressing blue wavelengths that experience less benefit from adaptive optics correction.

With the new planet candidate sizes we estimate the probability each has a radius $R > 15 R_\Earth$, the rough position of the boundary between gas giants and late-type stars. We then assume that every star in a blended KOI is an equally likely host for the transit, and measure an overall P$(R > 15 R_\Earth)$ as the mean average of probabilities for all possible hosts. This identifies potential false positives but does not constitute a full false positive calculation. Table~\ref{FPprobs} shows the results for all candidates for with P$(R > 15 R_\Earth) \geq 0.01$. We did not take into account the relative prevalence of planetary bodies and brown dwarfs, which might indicate that planets are generally more likely system members and invalidate the assumption that all possible configurations are equally likely.

Three candidates have been identified elsewhere as false positives. Likely the largest transiting object, KOI1447.01 appears in the \textit{Kepler} Eclipsing Binary Catalog \citep{slawson2011}. KOI0190.01 has a disposition of FALSE POSITIVE in the Exoplanet Archive from radial velocity measurements. KOI1112.01 has been identified as a false positive via ephemeris matching with the nearby KOI4720 by \citet{coughlin2014}, which notes that two stars are separated by only 4\farcs8, and states that the transit host is believed to be a third then-unobserved object. This implies the host is KOI1112B, which elevates our estimate to P$(R > 15 R_\Earth) = 0.95$.

\section{Discussion}\label{discussion}
Probabilities and uncertainties in this section are computed binomially by the method described in \citet{burgasser2003}.

Of the 93 companions with sufficient photometry, 13 (or $14.0\%^{+4.4\%}_{-2.9\%}$) are inconsistent with physical association with their primaries via KH07, while the B11 method gives 10 unassociated companions out of 53 examined (or $18.9\%^{+6.5\%}_{-4.2\%}$). All others are $<5\sigma$ consistent with a bound interpretation. Simulations have previously demonstrated that the vast majority (96\%) of narrowly-separated companions ($<1.0\arcsec$) are physically associated \citep{horch2014}. As 6 out of 40 (or $15\%^{+7.3\%}_{-4.0\%}$) narrowly-separated primary/companion pairs with fit results are inconsistent with a bound interpretation to $5\sigma$, our results are inconsistent with the Horch prediction to $\sim2.3\sigma$ and we see no evidence that narrowly separated ($<1.0\arcsec$) companions are more likely to be physically associated than KOI companions in general or than widely-separated companions, for which we determine 7 of 53 or $13.2\%^{+6.0\%}_{-3.3\%}$ are unbound.

Via transit reinterpretation we have 38 potential non-planetary objects out of 88 reinterpreted transiting objects. By summing the computed P$(R > 15 R_\Earth)$ values this sample has a mean of $12.8^{+3.5}_{-3.1}$, or $14.5\%^{+4.0\%}_{-3.5\%}$ of candidates with $(R > 15 R_\Earth)$. Considered with the previously reported $17.6\% \pm 1.5\%$ nearby-star (companion) probability of \citet{baranec2016}, we estimate a $(R > 15 R_\Earth)$ rate due to unresolved companions to be $2.6\% \pm 0.4\%$. This is a rough measurement of false positives in the KOI catalog and is easily consistent with the broad $<10\%$ false positive rate predicted by modeling \citep{morton2011}.

On the whole, the derived planet candidate sizes are only slightly larger than estimates from \citet{law2014}, henceforth L14. The exceptions are primarily those candidates listed in Table~\ref{FPprobs}. These are much larger than the L14 predictions as our analysis includes new sizes for the host stars and accounts for the change in distance estimates, whereas L14 used the original \textit{Kepler} predictions derived from blended light.

\subsection{KOI0191: Possible Coincident Multiple}
L14 noted that this system is \textit{a priori} unusual as the only multi-candidate KOI to have a large Jupiter-class candidate ($>10R_\Earth$) in a very close orbit ($P < 20$d). Assuming binarity (of KOI0191A/B), L14 calculated a planetary candidate size of 11.3$R_\Earth$ for the A scenario and 29.3$R_\Earth$ for B, making the potential KOI0191B/KOI0191.01 system a close eclipsing binary in a hierarchical triple. With the inclusion of \textit{JHK} photometry, we revise these estimates upward to 13.9$R_\Earth$ and 55.9$R_\Earth$, respectively. 

Although hot Jupiters were previously thought inconsistent with other short-period planets, the discovery of multiple planets in the WASP-47 system proves the arrangement does exist in nature \citep{becker2015}. Thus, we can not rule out that all four candidates are hosted by KOI0191A and are then planets.

\subsection{KOI0268: No Longer Habitable}
Both companions of KOI0268 have also been reported in \citet{adams2012}. KOI0268.01 was originally identified as a potentially habitable super-Earth with a radius of 1.7 $R_\Earth$ and equilibrium temperature of 295 K. L14 reports both companions, and notes that if the planet orbits either of them rather than the target A, the equilibrium temperature of the planet will probably not be in the habitable range. The candidate's equilibrium temperature in literature has since been revised upward from 295K to 470K as reported in the Exoplanet Archive. Our fitted stellar types (from KH07) yield an uninhabitable surface temperature of 650K if hosted by A. For B and C both we estimate a surface temperature of $\sim$350K, marginally allowing for the presence of liquid water, but the reinterpreted sizes imply a gas giant hosted by B or an eclipsing binary at C. Exomoons notwithstanding, this rules out habitability for KOI0268.01.

\subsection{KOI1447: Double Eclipsing Binary}
Both KOI1447.01 and KOI1447.02 are likely too large to be planets for either potential host, and .01 was included in the second release of the Kepler Eclipsing Binary Stars catalog \citep{slawson2011}. Given the size of both candidates and their short orbital periods (40.2d and 2.3d, respectively), it seems likely they would conflict with each other if in the same system. KOI1447 is then a unique double false positive, consisting of two merely visually associated eclipsing binaries with coincidentally low inclination.

\section{Conclusions}\label{conclusions}
We have obtained visible and near-infrared multi-wavelength photometry of 104 blended companions to 84 KOIs, validating the original detections by Robo-AO. We report additional companions not originally detected by Robo-AO's original investigation. We find that $14.5\%^{+3.8\%}_{-3.4\%}$ of the investigated companions are physically unassociated with their KOI primaries at the $5\sigma$ level. Additional follow-up is recommended to confirm this result, with spectroscopy of both targets the best means of measuring log $g$ to confirm actual sizes and distances, and to provide improved constraints on transit candidate size. We also find no evidence that narrowly separated KOI companions are more likely to be physically associated than widely separated companions, contrary to prior modeling work.

We have also reinterpreted 88 transit candidates, refining estimates of size given possible hosts and identifying 43 candidates potentially too large for planetary interpretation. With some assumptions, this produces an overall P$(R > 15 R_\earth)$ for transits with detected contaminating companions of $17.5\%^{+4.1\%}_{-3.7\%}$, or an overall P$(R > 15 R_\earth)$ for all KOIs (as a result of undetected companions) of $2.5\% \pm 0.4\%$. A more complete set of \textit{JHK} follow-up on KOI companions would refine this result.

Given the termination of \textit{Kepler}'s primary mission, solving host ambiguity for individual transit candidates is difficult. A close review of extant \textit{Kepler} data for astrometric motion or light curve re-analysis may detect centroid motion correlated with transit that would identify the host star. Independent investigations like radial velocity and ground-based AO transit imaging are possible but difficult and limited to bright targets and deep transits, respectively.

\section*{Acknowledgments}
D.A. is supported by a NASA Space Technology Research Fellowship, grant NNX13AL75H. The authors thank the NSTRF staff and the Office of the Chief Technologist for their assistance. C.B. acknowledges support from the Alfred P. Sloan Foundation.

This work was partially supported by the NASA XRP grant NNX15AC91G.

The Robo-AO system was developed by collaborating partner institutions, the California Institute of Technology and the Inter-University Centre for Astronomy and Astrophysics, and supported by the National Science Foundation under grants AST-0906060, AST-0960343, and AST-1207891, the Mt. Cuba Astronomical Foundation and by a gift from Samuel Oschin.

Some of the data presented herein were obtained at the W.M. Keck Observatory, which is operated as a scientific partnership among the California Institute of Technology, the University of California and the National Aeronautics and Space Administration. The Observatory was made possible by the generous financial support of the W.M. Keck Foundation.

The authors wish to recognize and acknowledge the very significant cultural role and reverence that the summit of Maunakea has always had within the indigenous Hawaiian community.  We are most fortunate to have the opportunity to conduct observations from this mountain.

This paper includes data collected by the \textit{Kepler} mission. Funding for the \textit{Kepler} mission is provided by the NASA Science Mission directorate.

Some of the data presented in this paper were obtained from the Mikulski Archive for Space Telescopes (MAST). STScI is operated by the Association of Universities for Research in Astronomy, Inc., under NASA contract NAS5-26555. Support for MAST for non-HST data is provided by the NASA Office of Space Science via grant NNX09AF08G and by other grants and contracts.

{\it Facilities:} \facility{PO:1.5m (Robo-AO), Keck:II (NIRC2-LGS).}

\bibliographystyle{astron}
\bibliography{biblio}

\begin{longtable*}{ c c c c c c }
\setlength{\extrarowheight}{3pt}\\
\caption{Measured \textit{JHK} Contrasts}\label{contrasts}\\
	object	 & sep($\arcsec$) & ang($^{o}$) &$\Delta m_J$ (mag)&$\Delta m_H$ (mag)&$\Delta m_{K}$ (mag)\\
	\hline
	\endfirsthead
	object	 & sep($\arcsec$) & ang($^{o}$) &$\Delta m_J$ (mag)&$\Delta m_H$ (mag)&$\Delta m_{K}$ (mag)\\
	\hline
	\endhead
	\endfoot
	\endlastfoot
	0190B & 0.180 $\pm$ 0.010 & 109.4 $\pm$ 3.2 &                  &                  & 0.642 $\pm$0.137 \\
  0191B & 1.660 $\pm$ 0.002 &  96.6 $\pm$ 0.1 & 2.588 $\pm$0.057 & 2.615 $\pm$0.054 & 2.626 $\pm$0.055 \\
	0268B & 1.753 $\pm$ 0.003 & 267.6 $\pm$ 0.1 & 3.056 $\pm$0.059 & 2.654 $\pm$0.057 & 2.553 $\pm$0.056 \\
	0268C & 2.528 $\pm$ 0.007 & 310.2 $\pm$ 0.1 & 3.810 $\pm$0.118 & 3.353 $\pm$0.127 & 3.984 $\pm$0.145 \\
  0401B & 1.986 $\pm$ 0.002 & 270.0 $\pm$ 0.1 & 2.066 $\pm$0.059 &                  & 1.635 $\pm$0.055 \\
  0425B & 0.491 $\pm$ 0.001 & 343.4 $\pm$ 0.1 &                  &                  & 0.831 $\pm$0.054 \\
  0511B & 1.300 $\pm$ 0.002 & 123.4 $\pm$ 0.1 & 2.221 $\pm$0.058 & 1.817 $\pm$0.007 & 1.707 $\pm$0.008 \\
  0511C & 3.865 $\pm$ 0.005 & 348.6 $\pm$ 0.1 & 5.055 $\pm$0.122 & 4.493 $\pm$0.077 & 4.308 $\pm$0.069 \\
  0628B & 2.748 $\pm$ 0.002 & 238.9 $\pm$ 0.1 &                  &                  & 3.000 $\pm$0.058 \\
  0628C & 1.828 $\pm$ 0.003 & 311.5 $\pm$ 0.2 &                  &                  & 3.871 $\pm$0.057 \\
  0687B & 0.680 $\pm$ 0.003 &  13.4 $\pm$ 0.4 &                  &                  & 1.251 $\pm$0.054 \\
  0688B & 1.734 $\pm$ 0.001 & 141.8 $\pm$ 0.1 & 1.552 $\pm$0.060 &                  & 1.373 $\pm$0.056 \\
  0712B & 0.470 $\pm$ 0.002 & 174.2 $\pm$ 0.3 & 0.435 $\pm$0.055 &                  & 0.351 $\pm$0.056 \\
  0931B & 1.263 $\pm$ 0.002 & 177.7 $\pm$ 0.1 &                  &                  & 3.227 $\pm$0.063 \\
  0984B & 1.764 $\pm$ 0.005 & 221.3 $\pm$ 1.4 & 0.064 $\pm$0.058 & 0.050 $\pm$0.054 & 0.059 $\pm$0.056 \\
  0987B & 1.974 $\pm$ 0.002 & 225.7 $\pm$ 0.3 & 2.612 $\pm$0.077 & 2.381 $\pm$0.058 & 2.239 $\pm$0.055 \\
	1066B & 1.690 $\pm$ 0.002 & 231.3 $\pm$ 0.1 &                  &                  & 2.949 $\pm$0.070 \\
	1067B & 2.932 $\pm$ 0.005 & 142.6 $\pm$ 0.1 &                  &                  & 2.785 $\pm$0.106 \\
	1112B & 3.068 $\pm$ 0.005 & 172.2 $\pm$ 0.1 & 3.607 $\pm$0.138 & 2.956 $\pm$0.081 & 2.758 $\pm$0.070 \\
	1151B & 0.758 $\pm$ 0.002 & 307.5 $\pm$ 0.7 &                  & 2.554 $\pm$0.055 & 2.407 $\pm$0.055 \\
	1214B & 0.371 $\pm$ 0.029 & 136.3 $\pm$ 0.3 &                  & 2.584 $\pm$0.055 & 2.455 $\pm$0.055 \\
  1274B & 1.085 $\pm$ 0.001 & 242.0 $\pm$ 0.1 & 2.801 $\pm$0.056 &                  & 2.506 $\pm$0.055 \\
	1359B & 1.387 $\pm$ 0.003 & 331.6 $\pm$ 0.4 &                  &                  & 2.168 $\pm$0.057 \\
	1375B & 0.784 $\pm$ 0.001 & 270.0 $\pm$ 0.1 &                  & 3.303 $\pm$0.069 & 3.393 $\pm$0.065 \\
  1442B & 2.114 $\pm$ 0.006 &  70.8 $\pm$ 0.1 & 4.155 $\pm$0.065 & 3.802 $\pm$0.056 & 3.631 $\pm$0.055 \\
	1447B & 0.282 $\pm$ 0.001 & 212.0 $\pm$ 0.1 &                  &                  & 0.625 $\pm$0.061 \\
  1536B & 0.580 $\pm$ 0.001 &  97.9 $\pm$ 0.1 &                  & 4.262 $\pm$0.136 & 4.177 $\pm$0.112 \\
  1546B & 0.603 $\pm$ 0.002 &  89.5 $\pm$ 0.1 & 0.940 $\pm$0.055 & 0.784 $\pm$0.055 & 0.726 $\pm$0.054 \\
  1546C & 2.915 $\pm$ 0.001 &   4.0 $\pm$ 0.1 & 3.224 $\pm$0.059 & 3.021 $\pm$0.071 & 2.945 $\pm$0.081 \\
	1546D & 4.119 $\pm$ 0.011 & 164.7 $\pm$ 0.1 & 3.338 $\pm$0.073 & 3.253 $\pm$0.077 & 3.479 $\pm$0.058 \\
	1613B & 0.214 $\pm$ 0.004 & 185.5 $\pm$ 1.3 & 1.136 $\pm$0.055 & 0.996 $\pm$0.055 & 0.997 $\pm$0.055 \\
	1700B & 0.274 $\pm$ 0.048 & 288.1 $\pm$10.7 &                  &                  & 0.551 $\pm$0.055 \\
	1784B & 0.278 $\pm$ 0.001 & 291.1 $\pm$ 0.1 &                  &                  & 0.781 $\pm$0.058 \\
  1845B & 1.999 $\pm$ 0.011 &  78.9 $\pm$ 0.5 & 3.238 $\pm$0.055 &                  & 2.886 $\pm$0.055 \\
  1845C & 2.958 $\pm$ 0.025 & 348.0 $\pm$ 0.2 & 4.264 $\pm$0.069 &                  & 4.400 $\pm$0.092 \\
  1880B & 1.713 $\pm$ 0.001 & 100.9 $\pm$ 0.1 & 3.936 $\pm$0.058 & 4.149 $\pm$0.057 & 4.282 $\pm$0.058 \\
  1884B & 0.934 $\pm$ 0.001 &  95.6 $\pm$ 0.1 & 2.642 $\pm$0.056 & 2.410 $\pm$0.056 & 2.305 $\pm$0.055 \\
  1884C & 1.838 $\pm$ 0.001 &  81.9 $\pm$ 0.1 & 3.075 $\pm$0.056 & 2.867 $\pm$0.057 & 2.731 $\pm$0.055 \\
  1884D & 2.567 $\pm$ 0.002 & 327.5 $\pm$ 0.1 & 3.590 $\pm$0.164 & 3.536 $\pm$0.141 & 3.204 $\pm$0.141 \\
  1891B & 2.066 $\pm$ 0.003 & 211.4 $\pm$ 0.1 & 4.340 $\pm$0.077 & 4.561 $\pm$0.060 & 4.596 $\pm$0.066 \\
  1916B & 0.252 $\pm$ 0.001 & 146.3 $\pm$ 0.1 & 1.201 $\pm$0.056 &                  & 1.054 $\pm$0.055 \\
  1979B & 0.842 $\pm$ 0.002 & 193.4 $\pm$ 0.1 & 2.291 $\pm$0.055 &                  & 1.822 $\pm$0.055 \\
	1989B & 0.816 $\pm$ 0.001 &  39.5 $\pm$ 0.1 &                  &                  & 2.921 $\pm$0.055 \\
  2001B & 1.167 $\pm$ 0.001 & 342.0 $\pm$ 0.1 &                  &                  & 4.320 $\pm$0.060 \\
  2009B & 1.513 $\pm$ 0.004 & 178.0 $\pm$ 0.1 & 3.042 $\pm$0.092 & 2.950 $\pm$0.061 & 2.750 $\pm$0.055 \\
  2059B & 0.394 $\pm$ 0.001 & 290.0 $\pm$ 0.1 &                  &                  & 0.539 $\pm$0.151 \\
	2069B & 1.128 $\pm$ 0.001 & 107.0 $\pm$ 0.1 &                  &                  & 3.195 $\pm$0.059 \\
  2083B & 0.255 $\pm$ 0.002 & 166.1 $\pm$ 0.3 &                  & 1.687 $\pm$0.056 & 1.600 $\pm$0.054 \\
	2117B & 0.334 $\pm$ 0.001 & 111.5 $\pm$ 0.1 &                  &                  & 0.531 $\pm$0.055 \\
  2143B & 2.184 $\pm$ 0.005 & 317.4 $\pm$ 0.1 & 3.200 $\pm$0.120 &                  & 3.457 $\pm$0.087 \\
  2159B & 2.009 $\pm$ 0.001 & 323.8 $\pm$ 0.1 &                  & 2.638 $\pm$0.063 & 2.476 $\pm$0.060 \\
	2247B & 1.917 $\pm$ 0.002 & 350.3 $\pm$ 0.1 &                  &                  & 3.867 $\pm$0.068 \\
  2289B & 0.948 $\pm$ 0.001 & 221.7 $\pm$ 0.1 &                  &                  & 2.938 $\pm$0.055 \\
	2317B & 1.512 $\pm$ 0.002 & 112.2 $\pm$ 0.1 &                  &                  & 3.923 $\pm$0.058 \\
  2363B & 1.952 $\pm$ 0.001 & 357.3 $\pm$ 0.1 &                  &                  & 5.041 $\pm$0.086 \\
  2377B & 2.185 $\pm$ 0.002 & 335.2 $\pm$ 0.1 & 0.828 $\pm$0.080 & 0.671 $\pm$0.073 & 0.629 $\pm$0.068 \\
  2377C & 3.903 $\pm$ 0.008 & 315.9 $\pm$ 0.1 & 3.925 $\pm$0.193 & 3.816 $\pm$0.146 & 3.551 $\pm$0.170 \\
  2377D & 2.540 $\pm$ 0.002 &  41.5 $\pm$ 0.1 & 4.234 $\pm$0.096 & 4.029 $\pm$0.116 & 3.752 $\pm$0.117 \\
	2413B & 0.308 $\pm$ 0.036 & 250.1 $\pm$ 8.7 &                  & 0.470 $\pm$0.109 & 0.170 $\pm$0.059 \\
  2443B & 1.384 $\pm$ 0.002 & 164.0 $\pm$ 0.1 & 4.133 $\pm$0.066 &                  & 3.632 $\pm$0.060 \\
  2542B & 0.769 $\pm$ 0.002 &  29.1 $\pm$ 0.2 & 0.896 $\pm$0.055 &                  & 0.602 $\pm$0.054 \\
	2554B & 0.372 $\pm$ 0.010 & 149.3 $\pm$ 1.6 &                  &                  & 0.267 $\pm$0.054 \\
	2554C & 3.547 $\pm$ 0.005 & 203.6 $\pm$ 0.1 &                  &                  & 2.960 $\pm$0.098 \\
	2601B & 1.598 $\pm$ 0.002 &  14.1 $\pm$ 0.1 &                  &                  & 0.966 $\pm$0.057 \\
	2601C & 1.480 $\pm$ 0.002 & 295.2 $\pm$ 0.1 &                  &                  & 2.979 $\pm$0.057 \\
	2601D & 3.059 $\pm$ 0.003 &  30.1 $\pm$ 0.2 &                  &                  & 4.899 $\pm$0.135 \\
	2657B & 0.744 $\pm$ 0.365 & 131.7 $\pm$ 1.8 & 0.145 $\pm$0.055 & 0.126 $\pm$0.055 & 0.106 $\pm$0.054 \\
	2664B & 1.190 $\pm$ 0.005 &  90.5 $\pm$ 0.2 &                  &                  & 1.103 $\pm$0.055 \\
	2681B & 1.132 $\pm$ 0.005 & 148.0 $\pm$ 0.3 &                  &                  & 0.431 $\pm$0.056 \\
  2705B & 1.900 $\pm$ 0.003 & 304.3 $\pm$ 0.2 & 2.565 $\pm$0.097 & 2.672 $\pm$0.099 & 2.584 $\pm$0.067 \\
  2711B & 0.472 $\pm$ 0.006 & 148.9 $\pm$ 0.2 & 0.149 $\pm$0.055 & 0.122 $\pm$0.055 & 0.118 $\pm$0.055 \\
  2722B & 3.224 $\pm$ 0.001 & 283.3 $\pm$ 0.2 &	                 & 3.937 $\pm$0.084 & 3.770 $\pm$0.064 \\
	2779B & 0.965 $\pm$ 0.010 &  66.5 $\pm$ 0.6 &                  &                  & 1.752 $\pm$0.055 \\
  2813B & 1.062 $\pm$ 0.001 & 261.1 $\pm$ 0.1 &                  &                  & 1.842 $\pm$0.055 \\
	2813C & 1.842 $\pm$ 0.005 & 187.8 $\pm$ 0.2 &                  &                  & 6.547 $\pm$ 0.237\\
	2837B & 0.355 $\pm$ 0.018 & 140.5 $\pm$ 2.8 & 0.218 $\pm$0.056 & 0.199 $\pm$0.055 & 0.200 $\pm$0.055 \\
	2859B & 0.454 $\pm$ 0.001 & 290.9 $\pm$ 0.1 & 3.262 $\pm$0.067 & 3.138 $\pm$0.066 & 2.890 $\pm$0.059 \\
  2869B & 1.625 $\pm$ 0.001 & 205.2 $\pm$ 0.1 &                  &                  & 5.670 $\pm$0.074 \\
  2904B & 0.699 $\pm$ 0.001 & 225.6 $\pm$ 0.1 & 2.705 $\pm$0.055 & 2.501 $\pm$0.055 & 2.446 $\pm$0.054 \\
	2971B & 0.300 $\pm$ 0.001 & 273.9 $\pm$ 0.1 & 4.503 $\pm$0.130 &                  & 3.568 $\pm$0.057 \\
	2971C & 3.561 $\pm$ 0.004 &  37.7 $\pm$ 0.1 & 7.656 $\pm$0.219 &                  & 5.931 $\pm$0.170 \\
  3020B & 0.379 $\pm$ 0.001 & 271.6 $\pm$ 0.1 &                  &                  & 1.266 $\pm$0.057 \\
  3020C & 3.862 $\pm$ 0.001 & 231.3 $\pm$ 0.1 &                  &                  & 5.008 $\pm$0.069 \\
  3029B & 0.251 $\pm$ 0.010 & 264.3 $\pm$ 2.3 &                  &                  & 0.135 $\pm$0.060 \\
  3029C & 2.543 $\pm$ 0.005 &   4.1 $\pm$ 0.1 &                  &                  & 3.440 $\pm$0.068 \\
  3029D & 1.734 $\pm$ 0.005 & 356.2 $\pm$ 0.2 &                  &                  & 4.489 $\pm$0.071 \\
  3069B & 1.790 $\pm$ 0.002 & 108.3 $\pm$ 0.1 & 1.579 $\pm$0.056 & 1.310 $\pm$0.055 & 1.265 $\pm$0.056 \\
  3106B & 0.272 $\pm$ 0.010 & 186.3 $\pm$ 2.2 &                  &                  & 1.221 $\pm$0.131 \\
  3377B & 0.265 $\pm$ 0.001 & 334.7 $\pm$ 0.1 &                  &                  & 0.485 $\pm$0.058 \\
  3377C & 1.406 $\pm$ 0.005 &  50.2 $\pm$ 0.2 &                  &                  & 3.741 $\pm$0.063 \\
	3401B & 0.648 $\pm$ 0.010 &  98.9 $\pm$ 0.9 &                  &                  & 1.877 $\pm$0.057 \\
  4004B & 1.954 $\pm$ 0.003 & 218.1 $\pm$ 0.1 &                  &                  & 2.373 $\pm$0.076 \\
  4209B & 0.976 $\pm$ 0.001 & 205.1 $\pm$ 0.1 & 0.322 $\pm$0.059 & 0.539 $\pm$0.056 & 0.570 $\pm$0.055 \\
  4292B & 1.950 $\pm$ 0.002 &  29.9 $\pm$ 0.1 &                  & 4.813 $\pm$0.075 & 4.542 $\pm$0.079 \\
  4331B & 0.335 $\pm$ 0.005 & 100.9 $\pm$ 1.0 &                  & 0.118 $\pm$0.055 & 0.125 $\pm$0.054 \\
  4407B & 2.453 $\pm$ 0.003 & 299.9 $\pm$ 0.1 & 2.286 $\pm$0.056 & 1.956 $\pm$0.056 & 1.893 $\pm$0.058 \\
  4407C & 2.660 $\pm$ 0.003 & 311.0 $\pm$ 0.1 & 4.479 $\pm$0.490 & 4.140 $\pm$0.712 & 4.654 $\pm$0.344 \\
  4463B & 2.457 $\pm$ 0.003 & 323.9 $\pm$ 0.1 & 0.160 $\pm$0.102 & 0.242 $\pm$0.082 & 0.259 $\pm$0.068 \\
  4634B & 0.281 $\pm$ 0.001 & 276.1 $\pm$ 0.1 &                  &                  & 0.653 $\pm$0.055 \\
  4768B & 1.339 $\pm$ 0.005 & 159.0 $\pm$ 0.2 &                  &                  & 2.608 $\pm$0.071 \\
  4822B & 0.559 $\pm$ 0.010 &  63.2 $\pm$ 1.0 &                  &                  & 4.503 $\pm$0.147 \\
  4871B & 0.922 $\pm$ 0.001 & 333.6 $\pm$ 0.1 & 3.126 $\pm$0.058 & 3.026 $\pm$0.057 & 3.038 $\pm$0.055 \\
	5578B & 0.322 $\pm$ 0.001 &  97.2 $\pm$ 0.1 &                  &                  & 1.681 $\pm$0.055 \\
  5762B & 0.221 $\pm$ 0.010 & 100.3 $\pm$ 2.5 &                  &                  & 0.833 $\pm$0.076 \\
\caption*{Relative locations and NIR contrast measurements of observed \textit{Kepler} Objects of Interest. Contrast uncertainties are systematically measured by varying the photometric aperture size. Use of co-added images reduces separation/angle measurement uncertainties to the single-pixel level.}\\
\end{longtable*}

\begin{longtable*}{ c c c c c c }
\setlength{\extrarowheight}{3pt}\\
\caption{Apparent Magnitudes of Resolved KOI Components}\label{magnitudes}\\
	KOI   &        $m_J$      &        $m_H$      &         $m_K$      &          $m_i$     &        $m_{Kep}$ \\
	\hline
	\endfirsthead
	KOI   &        $m_J$      &        $m_H$      &         $m_K$      &          $m_i$     &        $m_{Kep}$ \\
	\hline
	\endhead
	\endfoot
	\endlastfoot
0190A &         &         & 12.876 $\pm$0.053 &         & 14.419 $\pm$0.050\\ 
0190B &         &         & 13.517 $\pm$0.089 &         & 15.748 $\pm$0.137\\ 
0191A & 13.827 $\pm$0.023 & 13.419 $\pm$0.026 & 13.340 $\pm$0.037 &         & 15.057 $\pm$0.042\\ 
0191B & 16.414 $\pm$0.057 & 16.032 $\pm$0.056 & 15.961 $\pm$0.062 &         & 18.156 $\pm$0.453\\ 
0268A & 9.773  $\pm$0.021 & 9.609  $\pm$0.023 & 9.518  $\pm$0.019 &         & 10.599 $\pm$0.301\\ 
0268B & 12.826 $\pm$0.059 & 12.259 $\pm$0.056 & 12.074 $\pm$0.056 &         & 14.603 $\pm$0.317\\ 
0268C & 13.573 $\pm$0.117 & 12.963 $\pm$0.123 & 13.499 $\pm$0.144 &         & 16.199 $\pm$0.314\\ 
0401A & 12.845 $\pm$0.023 &         & 12.402 $\pm$0.027 &         & 14.076 $\pm$0.036\\ 
0401B & 14.909 $\pm$0.056 &         & 14.037 $\pm$0.051 &         & 16.975 $\pm$0.261\\ 
0425A &         &         & 13.571 $\pm$0.043 &         & 15.101 $\pm$0.043\\ 
0425B &         &         & 14.401 $\pm$0.056 &         & 15.957 $\pm$0.076\\ 
0511A & 13.222 $\pm$0.023 & 12.957 $\pm$0.035 & 12.883 $\pm$0.034 &         & 14.276 $\pm$0.036\\ 
0511B & 15.444 $\pm$0.059 & 14.816 $\pm$0.056 & 14.647 $\pm$0.055 &         & 17.387 $\pm$0.355\\ 
0511C & 18.280 $\pm$0.127 & 17.449 $\pm$0.084 & 17.194 $\pm$0.077 &         & 20.649 $\pm$0.383\\ 
0628A &         &         & 12.484 $\pm$0.025 & 13.773 $\pm$0.020 &         \\ 
0628B &         &         & 15.486 $\pm$0.061 & 18.287 $\pm$0.138 &         \\ 
0628C &         &         & 16.356 $\pm$0.059 & 18.591 $\pm$0.217 &         \\ 
0687A &         &         & 12.415 $\pm$0.026 & 13.771 $\pm$0.042 &         \\ 
0687B &         &         & 13.667 $\pm$0.046 & 15.808 $\pm$0.238 &         \\ 
0688A & 13.158 $\pm$0.028 &         & 12.858 $\pm$0.023 &         & 14.129 $\pm$0.044\\ 
0688B & 14.708 $\pm$0.054 &         & 14.231 $\pm$0.049 &         & 16.316 $\pm$0.244\\ 
0712A & 13.176 $\pm$0.032 &         & 12.771 $\pm$0.030 & 13.831 $\pm$0.072 &         \\ 
0712B & 13.615 $\pm$0.040 &         & 13.123 $\pm$0.037 & 15.010 $\pm$0.207 &         \\ 
0931A &         &         & 13.828 $\pm$0.049 &         & 15.318 $\pm$0.030\\ 
0931B &         &         & 17.055 $\pm$0.079 &         & 18.719 $\pm$0.136\\ 
0984A & 11.178 $\pm$0.037 & 10.823 $\pm$0.036 & 10.741 $\pm$0.034 & 12.100 $\pm$0.072 & 12.340 $\pm$0.049\\ 
0984B & 11.242 $\pm$0.036 & 10.873 $\pm$0.036 & 10.797 $\pm$0.036 & 12.113 $\pm$0.074 & 12.430 $\pm$0.049\\ 
0987A & 11.328 $\pm$0.021 & 10.951 $\pm$0.020 & 10.906 $\pm$0.013 & 12.356 $\pm$0.027 & 12.590 $\pm$0.030\\ 
0987B & 13.940 $\pm$0.074 & 13.331 $\pm$0.055 & 13.146 $\pm$0.050 & 16.423 $\pm$0.629 & 16.158 $\pm$0.103\\ 
1066A &         &         & 13.922 $\pm$0.040 &         & 15.647 $\pm$0.030\\ 
1066B &         &         & 16.950 $\pm$0.077 &         & 19.837 $\pm$0.184\\ 
1067A &         &         & 13.313 $\pm$0.035 &         & 14.710 $\pm$0.029\\ 
1067B &         &         & 16.095 $\pm$0.105 &         & 18.762 $\pm$0.157\\ 
1112A & 13.509 $\pm$0.050 & 13.251 $\pm$0.039 & 13.155 $\pm$0.049 &         & 14.650 $\pm$0.029\\ 
1112B & 17.122 $\pm$0.146 & 16.204 $\pm$0.085 & 15.913 $\pm$0.081 &         & 19.222 $\pm$0.057\\ 
1151A &         & 11.917 $\pm$0.020 & 11.857 $\pm$0.021 & 13.249 $\pm$0.035 &         \\ 
1151B &         & 14.474 $\pm$0.053 & 14.263 $\pm$0.054 & 16.716 $\pm$0.562 &         \\ 
1214A &         & 13.092 $\pm$0.027 & 12.978 $\pm$0.028 &         & 14.933 $\pm$0.051\\ 
1214B &         & 15.678 $\pm$0.054 & 15.431 $\pm$0.054 &         & 16.139 $\pm$0.134\\ 
1274A & 12.088 $\pm$0.020 &         & 11.638 $\pm$0.017 & 13.142 $\pm$0.025 &         \\ 
1274B & 14.892 $\pm$0.055 &         & 14.143 $\pm$0.052 & 16.911 $\pm$0.429 &         \\ 
1375A &         & 12.310 $\pm$0.020 & 12.279 $\pm$0.018 & 13.554 $\pm$0.022 &         \\ 
1375B &         & 15.613 $\pm$0.070 & 15.668 $\pm$0.064 & 17.936 $\pm$0.506 &         \\ 
1442A & 11.354 $\pm$0.024 & 11.015 $\pm$0.028 & 10.947 $\pm$0.022 & 12.298 $\pm$0.020 &         \\ 
1442B & 15.507 $\pm$0.067 & 14.819 $\pm$0.060 & 14.581 $\pm$0.058 & 18.971 $\pm$0.563 &         \\ 
1447A &         &         & 12.292 $\pm$0.030 &         & 13.248 $\pm$0.039\\ 
1447B &         &         & 12.917 $\pm$0.043 &         & 15.288 $\pm$0.160\\ 
1536A &         & 11.405 $\pm$0.017 & 11.349 $\pm$0.018 & 12.550 $\pm$0.020 &         \\ 
1536B &         & 15.669 $\pm$0.135 & 15.532 $\pm$0.114 & 17.816 $\pm$0.570 &         \\ 
1546A & 13.783 $\pm$0.030 & 13.459 $\pm$0.029 & 13.373 $\pm$0.030 &         & 14.760 $\pm$0.095\\ 
1546B & 14.725 $\pm$0.047 & 14.245 $\pm$0.045 & 14.096 $\pm$0.045 &         & 16.318 $\pm$0.377\\ 
1546C & 17.121 $\pm$0.077 & 16.711 $\pm$0.079 & 16.851 $\pm$0.062 &         & 18.587 $\pm$0.530\\ 
1546D & 17.010 $\pm$0.064 & 16.479 $\pm$0.072 & 16.319 $\pm$0.085 &         & 18.615 $\pm$0.560\\ 
1613A & 10.915 $\pm$0.025 & 10.680 $\pm$0.027 & 10.647 $\pm$0.024 &    $\dagger$     &         \\ 
1613B & 12.049 $\pm$0.046 & 11.677 $\pm$0.048 & 11.643 $\pm$0.043 &    $\dagger$     &         \\ 
1700A &         &         & 12.890 $\pm$0.030 &         & 14.822 $\pm$0.079\\ 
1700B &         &         & 13.446 $\pm$0.040 &         & 15.893 $\pm$0.192\\ 
1784A &         &         & 12.533 $\pm$0.027 &         & 14.093 $\pm$0.055\\ 
1784B &         &         & 13.310 $\pm$0.043 &         & 14.674 $\pm$0.087\\ 
1845A & 12.881 $\pm$0.023 &         & 12.281 $\pm$0.021 &    *    &    *    \\ 
1845B & 16.119 $\pm$0.057 &         & 15.166 $\pm$0.053 &    *    &    *    \\ 
1845C & 17.146 $\pm$0.071 &         & 16.680 $\pm$0.092 &    *    &    *    \\ 
1880A & 12.293 $\pm$0.022 & 11.634 $\pm$0.018 & 11.474 $\pm$0.012 &         & 14.480 $\pm$0.033\\ 
1880B & 16.231 $\pm$0.059 & 15.785 $\pm$0.059 & 15.754 $\pm$0.058 &         & 18.146 $\pm$0.441\\ 
1884A & 14.196 $\pm$0.045 & 13.789 $\pm$0.057 & 13.738 $\pm$0.060 &         & 15.509 $\pm$0.035\\ 
1884B & 16.836 $\pm$0.067 & 16.200 $\pm$0.079 & 16.041 $\pm$0.078 &         & 19.617 $\pm$0.588\\ 
1884C & 17.268 $\pm$0.068 & 16.652 $\pm$0.079 & 16.468 $\pm$0.079 &         & 20.383 $\pm$0.436\\ 
1884D & 17.794 $\pm$0.165 & 17.321 $\pm$0.144 & 16.943 $\pm$0.146 &         & 21.075 $\pm$0.349\\ 
1891A & 13.879 $\pm$0.021 & 13.331 $\pm$0.028 & 13.274 $\pm$0.029 &         & 15.284 $\pm$0.031\\ 
1891B & 18.220 $\pm$0.077 & 17.895 $\pm$0.062 & 17.872 $\pm$0.068 &         & 19.545 $\pm$0.444\\ 
1916A & 12.797 $\pm$0.023 &         & 12.493 $\pm$0.026 &         & 13.684 $\pm$0.035\\ 
1916B & 14.002 $\pm$0.047 &         & 13.547 $\pm$0.046 &         & 16.421 $\pm$0.229\\ 
1979A & 11.941 $\pm$0.025 &         & 11.601 $\pm$0.014 & 12.845 $\pm$0.029 &         \\ 
1979B & 14.235 $\pm$0.057 &         & 13.423 $\pm$0.045 & 16.047 $\pm$0.366 &         \\ 
1989A &         &         & 11.841 $\pm$0.018 & 13.144 $\pm$0.024 & 13.372 $\pm$0.030\\ 
1989B &         &         & 14.766 $\pm$0.054 & 17.363 $\pm$0.535 & 16.866 $\pm$0.151\\ 
2001A &         &         & 11.144 $\pm$0.019 & 12.835 $\pm$0.020 & 13.135 $\pm$0.030\\ 
2001B &         &         & 15.463 $\pm$0.060 & 17.411 $\pm$0.258 & 17.617 $\pm$0.219\\ 
2009A & 12.714 $\pm$0.021 & 12.387 $\pm$0.021 & 12.333 $\pm$0.022 &         & 13.848 $\pm$0.033\\ 
2009B & 15.751 $\pm$0.089 & 15.335 $\pm$0.061 & 15.085 $\pm$0.055 &         & 17.929 $\pm$0.491\\ 
2059A &         &         & 11.182 $\pm$0.063 &         & 13.246 $\pm$0.048\\ 
2059B &         &         & 11.724 $\pm$0.098 &         & 14.339 $\pm$0.104\\ 
2069A &         &         & 12.231 $\pm$0.020 & 13.582 $\pm$0.020 & 13.777 $\pm$0.031\\ 
2069B &         &         & 15.422 $\pm$0.060 & 18.874 $\pm$0.508 & 18.026 $\pm$0.504\\ 
2083A &         & 12.263 $\pm$0.020 & 12.230 $\pm$0.024 & 13.446 $\pm$0.037 & 13.871 $\pm$0.056\\ 
2083B &         & 13.948 $\pm$0.050 & 13.827 $\pm$0.050 & 16.164 $\pm$0.359 & 14.907 $\pm$0.134\\ 
2117A &         &         & 13.516 $\pm$0.037 &         & 16.236 $\pm$0.032\\ 
2117B &         &         & 14.047 $\pm$0.046 &         & 16.567 $\pm$0.034\\ 
2143A & 12.924 $\pm$0.025 &         & 12.505 $\pm$0.026 &         & 14.145 $\pm$0.031\\ 
2143B & 16.122 $\pm$0.118 &         & 15.965 $\pm$0.088 &         & 17.654 $\pm$0.283\\ 
2159A &         & 12.098 $\pm$0.019 & 12.087 $\pm$0.021 & 13.322 $\pm$0.025 &         \\ 
2159B &         & 14.731 $\pm$0.058 & 14.563 $\pm$0.059 & 17.340 $\pm$0.512 &         \\ 
2247A &         &         & 12.046 $\pm$0.022 &         & 14.384 $\pm$0.029\\ 
2247B &         &         & 15.916 $\pm$0.069 &         & 19.508 $\pm$0.213\\ 
2289A &         &         & 12.075 $\pm$0.018 & 13.214 $\pm$0.021 & 13.374 $\pm$0.030\\ 
2289B &         &         & 15.012 $\pm$0.055 & 17.540 $\pm$0.285 & 18.008 $\pm$0.297\\ 
2317A &         &         & 12.704 $\pm$0.026 &         & 14.298 $\pm$0.031\\ 
2317B &         &         & 16.628 $\pm$0.063 &         & 19.227 $\pm$0.194\\ 
2363A &         &         & 12.360 $\pm$0.018 &         & 14.369 $\pm$0.031\\ 
2363B &         &         & 17.407 $\pm$0.085 &         & 20.945 $\pm$1.338\\ 
2377A & 13.673 $\pm$0.038 & 13.286 $\pm$0.045 & 13.245 $\pm$0.040 &    *    &    *    \\ 
2377B & 14.501 $\pm$0.063 & 13.958 $\pm$0.062 & 13.876 $\pm$0.057 &    *    &    *    \\ 
2377C & 17.598 $\pm$0.196 & 17.112 $\pm$0.148 & 16.785 $\pm$0.173 &    *    &    *    \\ 
2377D & 17.909 $\pm$0.105 & 17.318 $\pm$0.126 & 17.004 $\pm$0.120 &    *    &    *    \\ 
2413A &         & 13.352 $\pm$0.046 & 13.345 $\pm$0.041 &         & 15.236 $\pm$0.101\\ 
2413B &         & 13.820 $\pm$0.068 & 13.515 $\pm$0.044 &         & 17.342 $\pm$0.554\\ 
2443A & 12.933 $\pm$0.024 &         & 12.574 $\pm$0.022 &         & 13.995 $\pm$0.030\\ 
2443B & 17.070 $\pm$0.070 &         & 16.209 $\pm$0.063 &         & 19.395 $\pm$0.540\\ 
2542A & 13.253 $\pm$0.027 &         & 12.525 $\pm$0.034 &         & 15.841 $\pm$0.056\\ 
2542B & 14.147 $\pm$0.043 &         & 13.125 $\pm$0.043 &         & 17.037 $\pm$0.142\\ 
2554A &         &         & 13.708 $\pm$0.042 &    *    &    *    \\ 
2554B &         &         & 13.977 $\pm$0.049 &    *    &    *    \\ 
2554C &         &         & 16.667 $\pm$0.101 &    *    &    *    \\ 
2601A &         &         & 12.850 $\pm$0.032 &         & 14.222 $\pm$0.111\\ 
2601B &         &         & 13.816 $\pm$0.051 &         & 15.646 $\pm$0.384\\ 
2601C &         &         & 15.828 $\pm$0.065 &         & 18.129 $\pm$0.245\\ 
2601D &         &         & 17.752 $\pm$0.137 &         & 19.872 $\pm$1.153\\ 
2657A & 12.333 $\pm$0.032 & 11.978 $\pm$0.031 & 11.936 $\pm$0.031 &         & 13.497 $\pm$0.084\\ 
2657B & 12.477 $\pm$0.035 & 12.104 $\pm$0.033 & 12.041 $\pm$0.033 &         & 13.770 $\pm$0.103\\ 
2664A &         &         & 13.877 $\pm$0.041 &         & 16.065 $\pm$0.040\\ 
2664B &         &         & 14.979 $\pm$0.056 &         & 16.897 $\pm$0.065\\ 
2681A &         &         & 14.460 $\pm$0.057 &         & 16.295 $\pm$0.040\\ 
2681B &         &         & 14.890 $\pm$0.063 &         & 17.547 $\pm$0.091\\ 
2705A & 11.667 $\pm$0.025 & 11.016 $\pm$0.028 & 10.822 $\pm$0.024 &         & 14.765 $\pm$0.297\\ 
2705B & 14.232 $\pm$0.091 & 13.690 $\pm$0.094 & 13.404 $\pm$0.064 &         & 17.956 $\pm$0.328\\ 
2711A & 13.248 $\pm$0.033 & 12.982 $\pm$0.033 & 12.992 $\pm$0.033 &         & 14.337 $\pm$0.046\\ 
2711B & 13.397 $\pm$0.038 & 13.103 $\pm$0.035 & 13.111 $\pm$0.038 &         & 14.457 $\pm$0.052\\ 
2722A &         & 12.110 $\pm$0.023 & 12.026 $\pm$0.018 &         & 13.274 $\pm$0.029\\ 
2722B &         & 16.049 $\pm$0.082 & 15.795 $\pm$0.067 &         & 19.149 $\pm$0.138\\ 
2779A &         &         & 13.623 $\pm$0.039 &         & 15.040 $\pm$0.048\\ 
2779B &         &         & 15.376 $\pm$0.060 &         & 17.586 $\pm$0.332\\ 
2813A &         &         & 11.697 $\pm$0.020 &         & 13.951 $\pm$0.068\\ 
2813B &         &         & 13.540 $\pm$0.050 &         & 14.958 $\pm$0.155\\ 
2813C &         &         & 18.263 $\pm$0.244 &         & 22.013 $\pm$0.680\\ 
2837A & 12.997 $\pm$0.033 & 12.822 $\pm$0.031 & 12.800 $\pm$0.032 &         & 13.873 $\pm$0.035\\ 
2837B & 13.212 $\pm$0.036 & 13.022 $\pm$0.035 & 13.000 $\pm$0.036 &         & 14.102 $\pm$0.037\\ 
2859A & 12.589 $\pm$0.021 & 12.181 $\pm$0.016 & 12.125 $\pm$0.016 &         & 13.997 $\pm$0.044\\ 
2859B & 15.849 $\pm$0.067 & 15.323 $\pm$0.066 & 15.013 $\pm$0.056 &         & 16.118 $\pm$0.215\\ 
2869A &         &         & 12.367 $\pm$0.018 &         & 13.750 $\pm$0.029\\ 
2869B &         &         & 18.039 $\pm$0.074 &         & 21.629 $\pm$0.163\\ 
2904A & 11.766 $\pm$0.021 & 11.518 $\pm$0.022 & 11.467 $\pm$0.018 &         & 12.848 $\pm$0.045\\ 
2904B & 14.474 $\pm$0.053 & 14.017 $\pm$0.054 & 13.911 $\pm$0.051 &         & 14.833 $\pm$0.206\\ 
2971A & 11.792 $\pm$0.021 &         & 11.482 $\pm$0.016 &         & 12.769 $\pm$0.031\\ 
2971B & 16.289 $\pm$0.126 &         & 15.053 $\pm$0.056 &         & 16.840 $\pm$0.094\\ 
2971C & 19.444 $\pm$0.225 &         & 17.411 $\pm$0.167 &         & 20.654 $\pm$1.884\\ 
3020A &         &         & 12.323 $\pm$0.023 &         & 13.591 $\pm$0.031\\ 
3020B &         &         & 13.586 $\pm$0.046 &         & 16.818 $\pm$0.104\\ 
3020C &         &         & 17.328 $\pm$0.072 &         & 20.511 $\pm$0.335\\ 
3029A &         &         & 13.869 $\pm$0.040 &    *    &    *    \\ 
3029B &         &         & 14.004 $\pm$0.043 &    *    &    *    \\ 
3029C &         &         & 17.309 $\pm$0.076 &    *    &    *    \\ 
3029D &         &         & 18.358 $\pm$0.083 &    *    &    *    \\ 
3069A & 13.916 $\pm$0.033 & 13.561 $\pm$0.032 & 13.482 $\pm$0.037 &         & 15.049 $\pm$0.032\\ 
3069B & 15.496 $\pm$0.054 & 14.872 $\pm$0.052 & 14.747 $\pm$0.054 &         & 17.249 $\pm$0.065\\ 
3106A &         &         & 14.018 $\pm$0.052 &         & 15.858 $\pm$0.060\\ 
3106B &         &         & 15.233 $\pm$0.108 &         & 16.605 $\pm$0.111\\ 
3377A &         &         & 13.321 $\pm$0.034 &         & 15.355 $\pm$0.029\\ 
3377B &         &         & 13.805 $\pm$0.043 &         & 19.175 $\pm$0.190\\ 
3377C &         &         & 17.062 $\pm$0.070 &         & 21.181 $\pm$0.284\\ 
3401A &         &         & 12.909 $\pm$0.026 &         & 14.788 $\pm$0.069\\ 
3401B &         &         & 14.782 $\pm$0.054 &         & 15.679 $\pm$0.144\\ 
4004A &         &         & 11.221 $\pm$0.021 &         & 12.722 $\pm$0.031\\ 
4004B &         &         & 13.596 $\pm$0.069 &         & 16.732 $\pm$0.111\\ 
4209A & 15.098 $\pm$0.042 & 14.635 $\pm$0.060 & 14.496 $\pm$0.056 &         & 16.073 $\pm$0.095\\ 
4209B & 15.421 $\pm$0.047 & 15.172 $\pm$0.066 & 15.066 $\pm$0.063 &         & 18.759 $\pm$0.758\\ 
4292A &         & 11.340 $\pm$0.017 & 11.294 $\pm$0.011 &         & 12.897 $\pm$0.030\\ 
4292B &         & 16.152 $\pm$0.076 & 15.837 $\pm$0.075 &         & 20.932 $\pm$0.343\\ 
4331A &         & 12.697 $\pm$0.033 & 12.629 $\pm$0.032 &         & 13.869 $\pm$0.296\\ 
4331B &         & 12.816 $\pm$0.037 & 12.755 $\pm$0.035 &         & 14.118 $\pm$0.296\\ 
4407A & 10.292 $\pm$0.026 & 10.071 $\pm$0.029 & 9.9666 $\pm$0.018 &    *    &    *    \\ 
4407B & 12.579 $\pm$0.055 & 12.029 $\pm$0.057 & 11.859 $\pm$0.053 &    *    &    *    \\ 
4407C & 14.782 $\pm$0.475 & 14.180 $\pm$0.695 & 14.610 $\pm$0.341 &    *    &    *    \\ 
4463A & 13.575 $\pm$0.054 & 13.189 $\pm$0.046 & 13.175 $\pm$0.043 &         & 16.347 $\pm$0.033\\ 
4463B & 13.728 $\pm$0.059 & 13.428 $\pm$0.052 & 13.435 $\pm$0.048 &         & 16.356 $\pm$0.033\\ 
4634A &         &         & 12.725 $\pm$0.031 &         & 13.876 $\pm$0.080\\ 
4634B &         &         & 13.380 $\pm$0.042 &         & 15.753 $\pm$0.386\\ 
4768A &         &         & 13.886 $\pm$0.050 &         & 15.738 $\pm$0.029\\ 
4768B &         &         & 16.495 $\pm$0.265 &         & 19.726 $\pm$0.260\\ 
4822A &         &         & 12.062 $\pm$0.016 &         & 13.475 $\pm$0.030\\ 
4822B &         &         & 16.573 $\pm$0.144 &         & 20.183 $\pm$0.443\\ 
4871A & 12.125 $\pm$0.023 & 11.884 $\pm$0.022 & 11.843 $\pm$0.013 &         & 13.107 $\pm$0.032\\ 
4871B & 15.254 $\pm$0.060 & 14.908 $\pm$0.058 & 14.880 $\pm$0.054 &         & 16.225 $\pm$0.189\\ 
5578A &         &         & 9.7288 $\pm$0.018 &         & 11.281 $\pm$0.048\\ 
5578B &         &         & 11.410 $\pm$0.047 &         & 13.057 $\pm$0.193\\ 
5762A &         &         & 14.158 $\pm$0.056 &         & 16.115 $\pm$0.097\\ 
5762B &         &         & 14.993 $\pm$0.073 &         & 16.764 $\pm$0.169\\ 
\caption*{Apparent magnitudes of individual stars from contrast measurements and literature values of blended system. The great majority of JHK values are from the 2MASS catalog \citep{liebert1995}, while sources for $i$ and $Kepler$ are more varied. All values are as reported in the Exoplanet Archive, except $JHK$ for KOI0268, which is linked to a spurious 2MASS entry.
* indicates that not all companions were detected by Robo-AO. Without contrast measurements for all objects we can not accurately determine the apparent magnitudes.
$\dagger$ indicates that although a contrast measurement has been made, there is no blended measurement, and we can not determine the apparent magnitude.}
\end{longtable*}

\end{document}